%% file: main.tex
\documentclass{article}
\usepackage{microtype}
\usepackage{graphicx}
\usepackage{subfigure}
\usepackage{booktabs} 

\usepackage{hyperref}

\usepackage{algorithm}  
\usepackage{algorithmic}
\usepackage{amsmath}
\usepackage{amssymb}
\usepackage{mathtools}
\usepackage{amsthm}
\usepackage{bbm}
\usepackage{booktabs}
\usepackage{microtype}
\usepackage{graphicx}
\usepackage{threeparttable}
\usepackage{multirow}

\usepackage[accepted]{icml2022}
\icmltitlerunning{Protein Property Prediction via Retrieved Sequence Augmentation }

\begin{document}

\twocolumn[
\icmltitle{Retrieved Sequence Augmentation for Protein Representation Learning}


\begin{icmlauthorlist}
\icmlauthor{Chang Ma}{hk}
\icmlauthor{Haiteng Zhao}{pk}
\icmlauthor{Lin Zheng}{hk}
\icmlauthor{Jiayi Xin}{hk}
\icmlauthor{Qintong Li}{hk}
\icmlauthor{Lijun Wu}{ms}
\icmlauthor{Zhihong Deng}{pk}
\icmlauthor{Yang Lu}{uw}
\icmlauthor{Qi Liu}{hk}
\icmlauthor{Lingpeng Kong}{hk}
\end{icmlauthorlist}

\icmlaffiliation{hk}{Department of Computer Science, The University of Hong Kong}
\icmlaffiliation{pk}{School of Intelligence Science and Technology, Peking University}
\icmlaffiliation{ms}{Microsoft Research Asia}
\icmlaffiliation{uw}{Department of Computer Science, University of Waterloo}
\icmlcorrespondingauthor{Chang Ma}{changma@connect.hku.hk}
\icmlcorrespondingauthor{Lingpeng Kong}{lpk@cs.hku.hk}

\icmlkeywords{Machine Learning, ICML}

\vskip 0.3in
]
\printAffiliationsAndNotice{}

\begin{abstract}

\input{01_abstract}
\end{abstract}

\section{Introduction}
\input{02_introduction}

\section{Related Work}
\input{03_related_work}

\section{Problem Statement and Notations}

\input{04_preliminaries}

\section{MSA Transformer as a Retrieval Augmentation Method}
\input{06_framework.tex}

\section{Retrieval Sequence Augmentations}
\label{sec:rsa}
\input{05_methodology}

\section{Experiments}
\input{07_experiments}

\section{Conclusions and Future Work}
\input{08_conclusions.tex}


\bibliography{icml2022}
\bibliographystyle{icml2023}

\newpage
\appendix
\onecolumn

\input{appendix}

\end{document}

%% file: 01_abstract.tex
Protein language models have excelled in a variety of tasks, ranging from structure prediction to protein engineering. However, proteins are highly diverse in functions and structures, and current state-of-the-art models including the latest version of AlphaFold rely on Multiple Sequence Alignments (MSA) to feed in the evolutionary knowledge. Despite their success, heavy computational overheads, as well as the de novo and orphan proteins remain great challenges in protein representation learning. In this work, we show that MSA-augmented models inherently belong to retrieval-augmented methods. Motivated by this finding, we introduce \textbf{R}etrieved \textbf{S}equence \textbf{A}ugmentation (RSA) for protein representation learning without additional alignment or pre-processing. RSA links query protein sequences to a set of sequences with similar structures or properties in the database and combines these sequences for downstream prediction. We show that protein language models benefit from the retrieval enhancement on both structure prediction and property prediction tasks, with a 5\% improvement on MSA Transformer on average while being 373$\times$ faster. In addition, we show that our model can transfer to new protein domains better and outperforms MSA Transformer on de novo protein prediction. Our study fills a much-encountered gap in protein prediction and brings us a step closer to demystifying the domain knowledge needed to understand protein sequences. Code is available on \url{https://github.com/HKUNLP/RSA}.

%% file: 02_introduction.tex
Proteins are the basic yet intricate building blocks of life, performing a vast array of functions within organisms, including catalyzing metabolic reactions, DNA replication, responding to stimuli, providing structure to cells, and transporting molecules from one location to another~\citep{garrett2016biochemistry}. Central to the enigma of these building blocks is the complex knowledge of protein relationships in their sequences, structures, and functions, which is a consequence of the interplay between physics and evolution~\citep{sadowski2009sequence}. Experimental and theoretical efforts have been made to unveil the structures and functions of emergent proteins~\citep{korendovych2020novo, anishchenko2021novo}, yet few methods can keep pace with the rapid accumulation of sequences~\citep{roy2010tasser}.

Recently, protein language models~\citep{esm1b, lin2022language, AhmedElnaggar2021ProtTransTC,jumper2021highly} have achieved remarkable progress in predicting protein functions and structures from sequences. Protein language models create a distribution of amino acids that matches the co-occurrence probability in their natural state, thereby capturing structural and evolutionary knowledge. In these approaches, all protein knowledge is implicitly stored in the parameters, and the quality of the language model distribution is highly dependent on pre-training and parameter scale. For example, ESM-2~\citep{lin2022language} shows that evolutionary depth saturates at lower model scales, and scaling up to a model size of billions is inevitable for protein modeling. To this end, we study enhancing the prediction of language models with a simple retrieval-based augmentation. 

Previous work~\citep{UrvashiKhandelwal2019GeneralizationTM, AnirudhGoyal2022RetrievalAugmentedRL, KelvinGuu2020REALMRL, DingminWang2022AugmentingMP} in natural language processing and machine learning has demonstrated that introducing related input sequences can effectively introduce domain knowledge without excessive backbone parameter size. In protein learning, a similar approach Multiple Sequence Alignment (MSA) has been adopted to introduce evolutionary knowledge into models by augmenting input with aligned homologous sequences.  MSA has improved deep learning performance on various models~\citep{rao2021msa, jumper2021highly, marks2011protein, hong2022s}, yet its success is often attributed to 
the alignment process that highlights co-evolution -- especially the alignment process that is central to direct-coupling analysis methods~\citep{morcos2011direct, marks2011protein, HetunandanKamisetty2013AssessingTU}. The most common practice for constructing MSA~\citep{remmert2012hhblits, altschul1998iterated, johnson2010hidden} is to build a Hidden Markov Model (HMM) profile for the entire sequence space of databases and then iteratively search for homologous sequences. Despite efforts to accelerate MSA construction~\citep{remmert2012hhblits, deorowicz2016famsa, hauser2016mmseqs}, this process is notoriously slow -- it takes HHblits~\citep{remmert2012hhblits} 10 seconds to perform a single iteration search on Pfam with 64 CPUs -- and requires pre-computing of a HMM profile. 

\begin{figure}[t]
    \centering
    
    \includegraphics[width=0.8\linewidth]{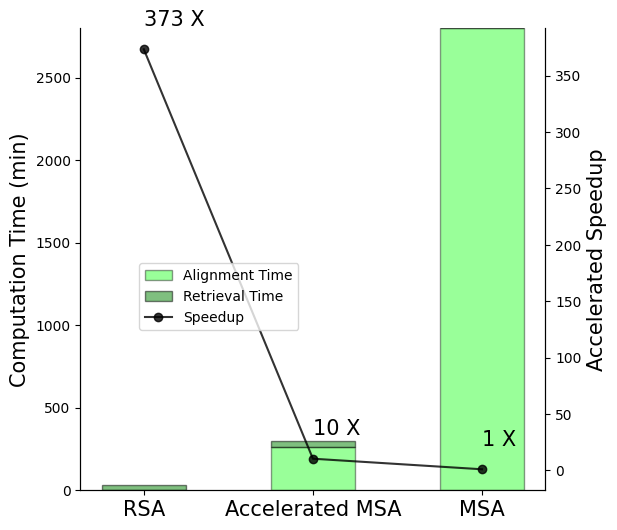}
        \caption{Illustration of speed up by RSA retrieval compared to MSA on secondary structure prediction dataset with 8678 sequences. Accelerated MSA refers to the MSA Transformer with MSA sequences retrieved by our RSA retriever. }
        \label{fig: speed compare}
\end{figure}

These considerations motivate us to rethink the role of MSA as a retrieval-based augmentation. Viewing MSA as a retrieval-augmentation method, it can be decomposed into two processes: retrieval and alignment. As shown in Figure~\ref{fig: speed compare}, the speed bottleneck of MSA is the alignment time, which is constrained by a quadratic complexity of $O(LD)$ ~\citep{remmert2012hhblits}, where $D$ is the database size, and $L$ is the protein length. Meanwhile, dense retrievers can be accelerated and use only a 100th of the time MSA needs to align a sequence ~\citep{hong2021fastmsa, johnson2019billion}. Moreover, the language of proteins encodes not only evolutionary knowledge but also other sources of information including structural and functional properties~\citep{xia2009micalign, o20043dcoffee}. Multiple sources of knowledge can be used to aid protein understanding when evolutionary knowledge is not available for orphan proteins and de novo (designed) proteins~\citep{perdigao2015unexpected, stefani2004protein, anishchenko2021novo}. Residue alignment imitates the mutation process in proteins, but empirically, present large language models have the potential to directly capture the evolutionary relationship between sequences without alignment information~\citep{riesselman2019accelerating}.

In light of these bottlenecks, We propose a simple yet effective \textbf{R}etrieved \textbf{S}equence \textbf{A}ugmentation (RSA) method as a general framework for augmenting protein sequences with related sequences from an unlabeled database. Specifically, RSA uses a pre-trained dense sequence retriever to retrieve protein sequences that are similar to the query sequence both in terms of homology as well as structure. These sequences are learned together with original input to help the model cover external knowledge and transfer to new domains. Extensive experiments on six tasks, including secondary structure prediction, contact prediction, homology prediction, stability prediction, subcellular localization, and protein-protein interaction demonstrate the effectiveness of our model. In addition, RSA overcomes the speed limit of MSA methods by directly inputting a batch of retrieved sequences into protein language models without performing the alignment process. Our main contributions are: 
\begin{itemize}
    \item Employing probabilistic analysis, we develop a unified framework that uses retrieval knowledge to enhance protein language models. Our theory along with our experiments strikes two novel perspectives: (1) MSA-augmented methods are essentially retrieval-augmented language models. Their performance can be explained by the injection of evolutionary knowledge. (2) The $O(N^2)$ complex alignment process is less necessary for deep protein language models. 
    
    \item We show that pre-trained dense retrievers can be faster and perform well in extracting homologous sequences and structurally similar sequences. 
    
    \item We leverage the retrieval augmentation framework to develop a new, fast method RSA. Unlike previous methods that combine protein language models with external knowledge, our method performs retrieval on-the-fly and requires no additional pre-training. We show that our model performs better than or competitively with  previous SOTAs. The result promises 
    new opportunities in using retrieval augmentation as a new paradigm in protein learning. Code and data are available in the supplementary material.

\end{itemize}

%% file: 03_related_work.tex
\textbf{Retrieval-Augmented Language Models}
The scaling laws of language models indicate that scaling up model size and training data are central to better performance~\citep {kaplan2020scaling}. However, larger language models are expensive to pre-train and may even be computationally heavy in inference. Retrieval-augmented language models~\citep{guu2020retrieval, he2021efficient, borgeaud2022improving} can achieve comparable performance on smaller models and are computationally more efficient by injecting external knowledge. Our RSA method is motivated by retrieval-augmented language models~\citep{guu2020retrieval, he2021efficient}, though we specifically focus on injecting protein knowledge and adapt the model for token-level tasks and better efficiency.

\textbf{Protein Language Models}
To model and further understand the protein sequence data, language models are introduced to train on mass data~\citep{heinzinger2019modeling, alley2019unified}. Large scale pre-training enables language models to learn structural and evolutionary knowledge~\citep{AhmedElnaggar2021ProtTransTC, jumper2021highly, lin2022language}. Despite these successes, many important applications still require MSAs and other external knowledge~\citep{rao2021msa,jumper2021highly,he2021pre,zhang2021co, ju2021copulanet, rao2020transformer}. MSAs have been shown effective in improving representation learning, despite being extremely slow and costly in computation. \citet{hu2022exploring} and \citet{hong2021fastmsa} use dense retrieval to accelerate multiple sequence augmentation, while still dependent on alignment procedures. Recent work~\citep{fang2022helixfold, lin2022language, wu2022high, chowdhury2022single} explores MSA-free language models though additional pre-training is involved. We take this step further to investigate retrieval-augmented protein language models that finds a balance between large scale pre-training and external knowledge. 

\input{table/01_unified_framework.tex}

%% file: table/01_unified_framework.tex
\begin{table*}[htbp]
\caption{Protein Retrieval Augmentation methods decomposed along a different axis. We formulate the aggregation function in the sequence classification setting and use a feed-forward neural network $\text{FFN}(\cdot)$ to map representations to logits. The proposed variants vary in design axis from the existing methods. $^\dagger$Note that MSA Transformer performs the aggregation in each layer of axial attention. 
}
\vspace{2mm}
\resizebox{\linewidth}{24mm}{
\begin{tabular}{lllcc}
\toprule
Method                           & Retriever Form       & Alignment Form   & Weight $\lambda_n$                              & Aggregation Function                                           \\ \midrule
\multicolumn{5}{c}{\textbf{Existing Methods}}                                                                                                                                                                                                           \\ 
Potts Model                      & MSA                  & Aligned          & —                                               & —                                                              \\
Co-evolution Aggregator          & MSA                  & Aligned          & $\frac{1}{N}$                                   & $\text{FFN}(\sum_{n=1}^N R_n(i) \lambda_n)$                    \\
MSA Transformer                  & MSA                  & Aligned          & $\sigma(\frac{XW_Q(R_nW_K)^T}{N\sqrt{d}})$ & $\text{FFN}(\sum_{n=1}^N R_n(i) \lambda_n)$$^\dagger$                    \\ \midrule
\multicolumn{5}{c}{\textbf{Proposed Variants}}                                                                                                                                                                                                          \\ 
Unaligned MSA Augmentation       & MSA                  & Not Aligned      & $\sigma( -||X-R_n||_2)$                         & $\sum_{n=1}^N \text{FNN}(R_n(i))\lambda_n$                     \\
Accelerated MSA Transformer      & Dense Retrieval      & Aligned          & $\sigma(\frac{XW_Q(R_nW_K)^T}{N\sqrt{d}})$ & $\text{FFN}(\sum_{n=1}^N R_n(i) \lambda_n)$                    \\
\multicolumn{1}{l}{Retrieval Sequence Augmentation} & \multicolumn{1}{l}{Dense Retrieval} & \multicolumn{1}{l}{Not Aligned} & \multicolumn{1}{l}{$\sigma(-||X - R_n||_2)$} & \multicolumn{1}{l}{$\sum_{n=1}^N \text{FFN}(\text{Embed}(x; r_n))\lambda_n$} \\ \bottomrule
\end{tabular}}

\label{table: pipeline}
\vspace{-2mm}
\end{table*}

%% file: 04_preliminaries.tex
The task of protein representation learning is to learn embeddings of protein sequences that can be transferred to downstream tasks with finetuning. For a protein $x$ with $L$ amino acids, it can be denoted as $x = [o_1, o_2,...o_L ]$, where each token $o_i$ denotes one of the 25 essential amino acids. We implement the embedding functions using BERT-style Transformer encoder $\text{{Embed}}( x) = [h_1, h_2, ...h_L]^T$, where $h_i \in \mathbbm{R}^d$ is a $d$-dimensional token representation for $o_i$. For token property prediction (i.e., secondary structure prediction), pairwise prediction (i.e., contact prediction), and sequence property prediction (i.e., protein engineering) tasks, the probabilities are obtained through pooling operations defined below:
\begin{align*}
&p(y_{\text{{Token}}}|o_i) = \text{{FFN}}(h_i), \\
    &p(y_{\text{{Pairwise}}}|o_i, o_j) = \text{{FFN}}([h_i ; h_j]), \\
    &p(y_{\text{{Sequence}}}|x) = \text{{FFN}}(\text{{Mean}}([h_1, h_2, ...h_L]).
\end{align*}

%% file: 06_framework.tex
In this section, we introduce a unified probabilistic framework to connect the MSA-based models with retrieval augmentations. We also offer a new holistic view on understanding these models, that is the retrieved protein sequences enhance the performance of pre-trained protein models by providing evolutionary knowledge in a similar  way as the MSA sequences do. 

Inspired by \citet{guu2020retrieval} and the probabilistic form of MSA Transformer,  we propose a general framework, \emph{protein retrieval augmentation}, that aims to unify several state-of-the-art evolution augmentation methods. Specifically, we consider these methods as learning a downstream predictor $p(y|x)$ based on an aggregation of homologous protein representations $R_{1...N}$.  
From the view of retrieval, $p(y|x)$ is decomposed into two steps: \emph{retrieve} and \emph{predict}. For a given input $x$, the retrieve step first finds possibly helpful protein sequence $r$ from a sequence corpus $\mathcal{R}$ and then predict the output $y$ conditioning on this retrieved sequence. We treat $r$ as a latent variable and in practice, we approximately marginalized it out with top-$N$ retrieved sequences:
\begin{align}
     p(y|x) = \sum_{r\in \mathcal{R}} p(y|x,r) p(r|x) \approx \sum_{n=1}^N p(y|x,r_n)p(r_n|x).
     \label{probality explanition}
\end{align} 
The probability $p(r|x)$ denotes the possibility that $r$ is sampled from the retriever given $x$. Intuitively it measures the similarity between the two sequences $r$ and $x$. This framework also applies to the MSA-based augmentation methods. We explain in detail using a state-of-the-art MSA-augmentation model \textit{MSA Transformer}~\citep{rao2021msa} as an example. In MSA Transformer, the layers calculate self-attention both row-wise and column-wise. Column-wise attention is defined as follows, given $W_Q$, $W_K$, $W_V$, $W_O$ as the parameters in a typical attention function:
\begin{align}
   R_s(i) = \sum_{n=1}^N \sigma(\frac{R_s(i)W_Q(R_n(i)W_K)^T}{N\sqrt{d}})R_n(i)W_VW_O,
   \label{msa formula}
\end{align}
where $R_n(i)$ denotes the $i$-th token representation of the $n$-th MSA sequence after performing the row-wise attention. Note that in MSA input, the first sequence $r_1$ is defined as the original sequence $x$. Then for a token prediction task, we define the $i$-th position  output as $y$ and the predicted distribution $p(y|x)$ can be expressed as: 
 \begin{equation}
    \begin{aligned}
        p(y|x) &=  \sum_{n=1}^N \sigma(\frac{R_1W_Q(R_nW_K)^T}{N\sqrt{d}})(R_n W_VW_OW_y) \\
    &= \sum_{n=1}^N p(y|x, r_n) \lambda_n 
    = \sum_{n=1}^N p(y|x, r_n)p(r_n|x),
    \label{eq: probabilistic look}
    \end{aligned}
\end{equation}
where $\lambda_n = \sigma(\frac{R_1(i)W_Q(R_n(i)W_K)^T}{N\sqrt{d}})$ is the weighting norm that represents the similarity of retrieved sequence $r_n$ and original sequence $x$;  $p(y|x, r_n)$ is a predictor that maps the row-attention representation of $r_n$ and $x$ to label. 

Eq.\ref{eq: probabilistic look} gives a retrieval-augmentation view of MSA Transformer that essentially retrieves homologous sequences with multiple sequence alignment and aggregates representations of homologous sequences with regard to their sequence similarity.
Taking one step further, we define a set of design dimensions to characterize the retrieving and aggregation processes. We detail the design dimensions below and illustrate how popular models (Appendix \ref{augmentation methods}) and our proposed methods (\S\ref{sec:rsa}) fall along them in Table \ref{table: pipeline}. These design choices includes:
\begin{itemize}
    \item \textbf{Retriever Form} indicates the retriever type used. Multiple Sequence Alignment is a discrete retrieval method that uses E-value thresholds~\citep{ye2006blast} to find homologous sequences. Dense retrieval~\citep{johnson2019billion} has been introduced to accelerate discrete sequence retrieval. The method represents the database with dense vectors and retrieves the sequences that have top-$k$ vector similarity with the query. 
    \item \textbf{Alignment Form} indicates whether retrieved sequences are aligned, as illustrated in Appendix Figure \ref{fig: illustrated msa}. 
    \item \textbf{Weight Form} is the aggregation weight of homologous sequences, as the $p(r_n|x)$ in Eq. \ref{eq: probabilistic look}. Here we denote this weight as $\lambda_n$. Traditionally, aggregation methods consider the similarity of different homologous sequences to be the same and use average weighting. MSA Transformer also use a weighted pooling method though the weights of $\lambda_n$ use global attention and are dependent on all homologous sequences. 
    \item \textbf{Aggregation Function} is how the representations of homologous sequences are aggregated to the original sequence to form downstream prediction, as in $p(y|x,r)$. For example, considering the sequence classification problem, a fully connected layer maps representations to logits. MSA Transformer first aggregates the representations $R_n$ and then maps the aggregated representation to logits $y$, and the retrieval augmentation probabilistic form first maps each representation to logits $p(y|x,r_n)$ and then linearly weight the logits with $\lambda_n$ in Eq. \ref{eq: probabilistic look}. 
\end{itemize}
Our discussion and formulation so far reach the conclusion that MSA augmentation methods intrinsically use the retrieval augmentation approach. This highlights the potential of RSA to replace MSA Augmentations as a computationally effective and more flexible method. 

However, MSA-based methods claim a few advantages: the \textit{alignment} process can help the model capture column-wise residue evolution; and the \textit{MSA Retriever} uses a discrete, token-wise search criterion that ensures all retrieved sequences are homology. We propose two novel variants to help verify these claims.

\paragraph{Unaligned MSA Augmentation.} \label{accelerated msa} MSA modeling traditionally depends on the structured alignment between sequences to learn evolutionary information. However, deep models have the potential to learn patterns from unaligned sequences. \citet{riesselman2019accelerating} shows that the mutation effect can be learned from unaligned sequences using autoregressive models. Therefore, we first introduce this variant that uses the homologous sequences from MSA to augment representations without alignment. 

\paragraph{Accelerated MSA Transformer.} This variant explores substituting the discrete retrieval process in MSA with a dense retriever. We use the K-nearest neighbor search to find the homologous sequences. We still align the sequences before input into MSA Transformer. We introduce this variant to find if MSA builder has an advantage over our pre-trained dense retriever in finding related sequences.

An empirical study of the performance of these models can be found in Subsection \ref{section: msa alignment experiments}.



%% file: 05_methodology.tex
\begin{figure}[htbp]
\includegraphics[width=\linewidth]{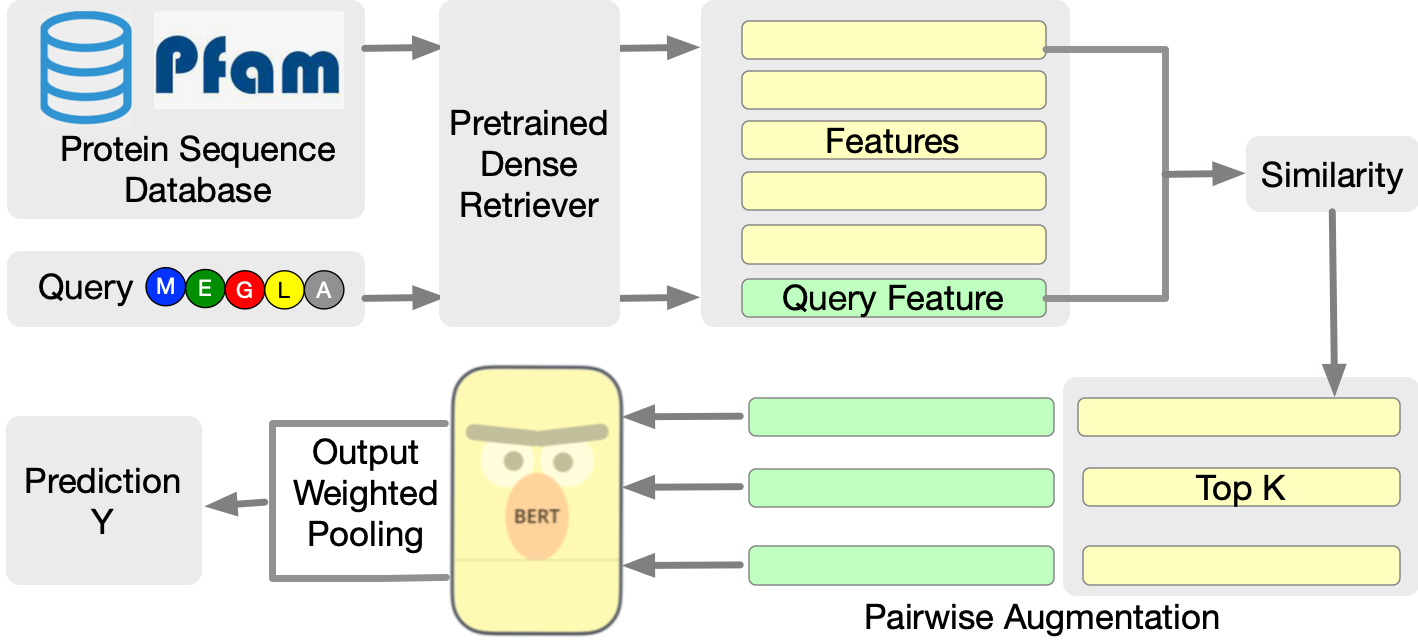}
    \caption{A brief overview of the proposed RSA protein encoding framework. Based on a query protein, RSA first retrieves related protein data from the database based on the top K similar features encoded by a pretrained retrieval model. Then we augment the query protein into pairs with each retrieved data and feed them into the protein model for protein tasks.} 
    \label{overview}
\end{figure}

\input{table/00_motivation_recall.tex}
Existing knowledge augmentation methods for protein representation learning are either designed for a specific task or require cumbersome data preprocessing. Motivated by the potential of pre-trained retrievers to identify proteins that are homologous or geometric similar, we propose a pipeline, RSA (\textbf{R}etrieval \textbf{S}equence \textbf{A}ugmentation), to directly augment protein models on-the-fly. 
Our model implementation follows the \textit{retrieve-then-predict} framework in Eq. \ref{probality explanition}. We elaborate on the model architecture implementations in Subsection \ref{section: 4.1} and describe model training in Subsection \ref{section 4.2}.

\input{table/03_main_result.tex}

\subsection{Model Architectures\label{section: 4.1}}
The RSA model comprises of a neural sequence retriever $p(r|x)$, and a protein model that combines both original input and retrieved sequence to obtain prediction $p(y|x, r)$. 

\subsubsection{RSA Retriever}
The retriever is defined as finding the sequences that are semantically close to the query. Denote retriever model as $G$ which encode protein sequence and output embeddings.
\begin{equation}
\begin{aligned}
    p(r|x) &= \frac{\exp f(x,r)}{\sum_{r'\in \mathcal{R}} \exp f(x,r')}, \\ f(x,r) &= -||G(x)-G(r)||_2
\end{aligned}
\label{pzx}
\end{equation}
 The similarity score $f(x,r)$ is defined as the negative L2 distance between the embedding of the two sequences. The distribution is the softmax distribution over similarity scores. 

For protein retrieval, we aim to retrieve protein sequences that have similar structures or are homologous to the query sequence. Motivated by the k-nearest neighbor retrieval experiment with ESM-1b \citep{esm1b} pre-trained embeddings (as shown in Table \ref{table: retrieve experiment} and 
Figure \ref{retrieval plot}), we implement the embedding functions using a 34-layer ESM-1b encoder. We obtain sequence embeddings by performing average pooling over token embeddings. Note that finding the most similar proteins from a large-scale sequence database is computationally heavy. To accelerate retrieval, we use Faiss indexing \citep{faiss}, which uses clustering of dense vectors and quantization to allow efficient similarity search at a massive scale.

\subsubsection{RSA Encoder}

\textbf{Retrieval Augmented Protein Encoder} 
Given a sequence $x$ and a retrieved sequence $r$ with length $L$ and $M$ respectively, the protein encoder combines $x$ and $r$ for prediction $p(y|x, r)$. To make our model applicable to any protein learning task, we need to augment both sequence-level representation and token-level representation. To achieve this, we concatenate the two sequences before input into the transformer encoder, which uses self-attention to aggregate global information from the retrieved sequence $r$ into each token representation. 
\begin{align*}
   & {A} = \sigma(\frac{(H_{[x;r]}W^Q)(H_{[x;r]}W^K)^T} {\sqrt{d}}),  A = [A_{x}; A_{r}] \\
   & Attn(H_{[x;r]}) = (A_xH_{x}W^V + A_rH_{r}W^V)W^O 
\end{align*}
where $H_{[x;r]} = [h_1^x, h_2^x, ..., h_L^x, h_1^r... h_M^r]^T$ denotes the input embedding of original and retrieved sequences. The output token representation $h_i$ automatically learns to select and combine the representation of retrieved tokens. This can also be considered a soft version of MSA alignment. After computing for each pair of $(x,r)$, we aggregate them by weight $p(r|x)$ defined in Eq. \ref{pzx}.

\subsection{RSA Training\label{section 4.2}}

\textbf{Training}
For downstream finetuning, we maximize $p(y|x)$ by performing training on the retrieval augmented protein encoder. We freeze the retriever parameters during training. For a query sequence with $N$ retrieved proteins, the computation cost is $N$ times the original model, $O(NL^2)$ for a transformer encoder layer, which is more efficient than the MSA Transformer with a $O(NL^2) + O(N^2L)$ computation cost. Also, the retrieval is performed on the fly.

%% file: table/00_motivation_recall.tex
\begin{table}[htbp]
\caption{Recall and Precision for retrieving top 100 protein sequences with ESM1b embeddings. In dataset Pfam and SCOPe, we test whether retrieved proteins are of the same Family, Superfamily, or Fold as query protein, and report the recall and precision.}
\vspace{2mm}
\centering
\resizebox{\linewidth}{12mm}{
\begin{tabular}{llll}
\toprule
Retrieval Task (Top 100)  & Type              &  Recall &  Precision \\
\midrule
Pfam  - Family           & Homology  & 100  & 90.42   \\
SCOPe - Fold            & Structural   &  100    & 65.98 \\  
SCOPe - Superfamily     & Structural   &  100    &  46.00\\  
SCOPe - Family          & Structural   &  100    &  24.71 \\  
\bottomrule
\end{tabular}}
\vspace{-3mm}
\label{table: retrieve experiment} 
\end{table}

%% file: table/03_main_result.tex
\begin{table*}[htbp]
\caption{Main Results for vanilla protein representation learning methods, knowledge-augmented baselines and our proposed RSA method. Note that \textit{italized} result is reported by corresponding related work. The last column reports average result on all six tasks. For MSA Transformer and RSA, we all use 16 sequences (N=16) for augmentation. For Gremlin Potts model, we use the full MSA.}
\vspace{2mm}
\label{Table: main result}
\resizebox{\linewidth}{!}{
\setlength{\tabcolsep}{1.0mm}{
\begin{tabular}{lcccccccccc}
\toprule
     Method    & Pretrain & Knowledge  & Knowledge                                                                     & SSP                & Contact        & Homology              & Stability  & Loc               & PPI    & Avg              \\
          & & Pretrain & Injection \\

     \midrule

 Transformer    & $\times$ &  $\times$  &  $\times$                                                      &         0.384             &      0.274              &        0.101           &    0.422     &   0.541               &       0.616  &      0.345      \\  
 LSTM  & $\times$ &  $\times$ &  $\times$  &         \textit{0.596}             &     \textit{0.263}               &        \textit{0.181}          &    \textit{0.591}     &   \textit{0.629}              &       \textit{0.638}     &      0.404      \\ \midrule

 RSA (Transformer backbone) & $\times$ &  $\times$ &   \checkmark    & \multicolumn{1}{c}{0.541} & \multicolumn{1}{c}{0.332} & \multicolumn{1}{c}{0.346} & \multicolumn{1}{c}{0.602}  & \multicolumn{1}{c}{0.591}& \multicolumn{1}{c}{0.700}&\multicolumn{1}{c}{0.518} \\                     
\midrule
 \midrule
 
 ESM-1b  & 
 \checkmark &   $\times$   &  $\times$                                                        &    \textit{0.716}                 &      \textit{0.458}                 &   \textit{0.978}                         &   \textit{0.695}        &           \textit{0.781}           &          \textit{0.782}       &   0.668  \\
 ProtBERT    & 
 \checkmark &   $\times$    &  $\times$                                                   &         0.691             &      0.556                &   0.528                   &          0.651            &         0.771            &    0.688      &     0.579       \\
 MSA Transformer (MSA N=1)    & 
 \checkmark &    \checkmark & $\times$                                                        &        0.594            &      0.397                &   0.880            &             0.767         &    0.668                 &      0.633       &    0.592        \\\midrule
 Gremlin~\citep{balakrishnan2011learning}  & 
 $\times$ & 
 $\times$ &   \checkmark  & —                    &       0.507               & —                      & —                    & —                    & —            & —        \\

 MSA Transformer        & 
 \checkmark &   \checkmark   &   \checkmark                              & \multicolumn{1}{c}{0.654} & \multicolumn{1}{c}{0.618} & \multicolumn{1}{c}{0.958}  & \multicolumn{1}{c}{\textbf{0.796}} & \multicolumn{1}{c}{0.694}& \multicolumn{1}{c}{0.751}& \multicolumn{1}{c}{0.672} \\

 OntoProtein~\citep{zhang2022ontoprotein}   & 
 \checkmark &   $\times$ &   \checkmark  & \textit{0.68}                   &      \textit{0.40}               &       0.96           & \textit{0.75}                   & —                    & —            & —        \\

 PMLM \citep{he2021pre}  & 
 \checkmark &   \checkmark &   $\times$ 
                             &       \textbf{\textit{0.728}}  & \textbf{\textit{0.717}}         &  \textit{0.946}                  &   —                    &    —                   &  —           &   —           \\ \midrule

 RSA (ProtBERT backbone)    & 
 \checkmark  &   $\times$ &   \checkmark            & \multicolumn{1}{c}{0.691} & \multicolumn{1}{c}{\textbf{0.717}} & \multicolumn{1}{c}{\textbf{0.987}}& \multicolumn{1}{c}{0.778} & \multicolumn{1}{c}{\textbf{0.795}} & \multicolumn{1}{c}{\textbf{0.827}}& \multicolumn{1}{c}{\textbf{0.723}} \\

\bottomrule
\end{tabular}}}
\vspace{-3mm}
\label{main result}
\end{table*}

%% file: 07_experiments.tex

\subsection{General Setup} \label{General Setup}
\textbf{Downstream tasks} In order to evaluate the performance of our trained model, six datasets are introduced, namely secondary structure prediction, contact prediction, remote homology prediction, subcellular localization prediction, stability prediction, and protein-protein interaction. Please refer to Appendix Table \ref{Table: table task} for more statistics of the datasets. The train-eval-test splits follow TAPE benchmark \citep{rao2019evaluating} for the first four tasks and PEER benchmark \citep{xu2022peer} for subcellular localization and protein-protein interaction. The introduction to datasets is in Appendix \ref{Introduction to the datasets}.

\textbf{Retriever and MSA Setup} Limited by available computation resources, we build a database on Pfam~\citep{pfam-gebali2018} sequences, which covers 77.2\% of the UniProtKB~\citep{apweiler2004uniprot} database and reaches the evolutionary scale. We generate ESM-1b pre-trained representations of 44 million sequences from Pfam-A and use Faiss~\citep{johnson2019billion}
to build the retrieval index. For a fair comparison, the MSA datasets are also built on the Pfam database. We use HHblits~\citep{remmert2012hhblits} to extract MSA. The details are shown in Appendix \ref{Introduction to the retrievers}.


\textbf{Baselines} We apply our retrieval method to both pre-trained and randomly initialized language models. Following \citet{rao2019evaluating} and \citet{rao2021msa}, we compare our model with vanilla protein representation models, including LSTM\citep{liu2017deep}, Transformers\citep{vaswani2017attention} and pre-trained models ESM-1b\citep{esm1b}, ProtBERT\citep{elnaggar2020prottrans}. We also compare with state-of-the-art knowledge-augmentation models: Potts Model\citep{balakrishnan2011learning}, MSA Transformer\citep{rao2021msa} that inject evolutionary knowledge through MSA, OntoProtein\citep{zhang2022ontoprotein} that uses gene ontology knowledge graph to augment protein representations and PMLM\citep{he2021pre} that uses pair-wise pretraining to improve co-evolution awareness. We use the reported results of LSTM from \citet{zhang2021co, xu2022peer}.

\textbf{Training and Evaluation} 
Our RSA model is applicable to any global-aware encoders. To demonstrate RSA as a general method, we perform experiments both with a shallow transformer encoder, and a large pre-trained ProtBERT encoder. The Transformer model has 512 dimensions and 6 layers. All self-reported models use the same truncation strategy and perform parameter searches on the learning rate, warm-up rate, seed, and batch size. For evaluation, we choose the best-performing model on the validation set and perform prediction on the test set. 


 \begin{figure}[htbp]
    \centering
    
    \includegraphics[width=0.5\linewidth]{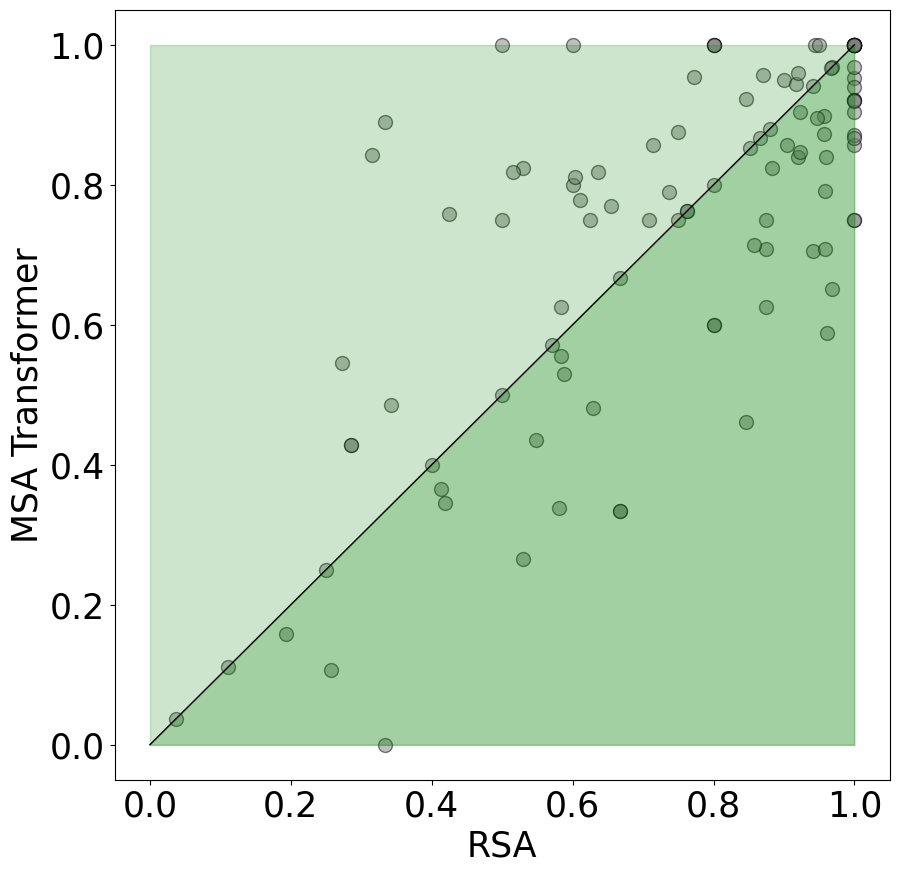}
        \caption{Contact Prediction of RSA and MSA Transformer on De Novo Proteins. We plot samples that RSA have better predictions under the diagonal line.}
        \label{fig: de novo contact}
        \vspace{-3mm}
\end{figure}

\subsection{Main Results}
\label{Main Results}

We show the result for downstream tasks in Table 
\ref{main result}, including models with/without pretraining, and with/without knowledge augmentations. We form the following conclusion: \textbf{Retrieval Sequence Augmentations perform on par with or even better than other knowledge-augmented methods without additional pre-training.} The last two blocks compare our method with previous augmentation methods. Our method outperforms MSA Transformer on average by 5\% and performs on par with PMLM on structure and evolution prediction tasks. Notably, both MSA Transformer and PMLM perform additional pre-training with augmentations, while our method uses no additional pre-training. From the results, we can see that RSA combined transformer model also improves by 10\% than other shallow models, demonstrating the effectiveness of our augmentation to both shallow models and pre-trained models.

\input{table/08_domain_adpation.tex}

\subsection{Retrieval Augmentation for Domain Adaptation}

{We investigate the model's transfer performance in domains with distribution shifts. We train our model on the Remote Homology dataset, and test it on three testsets with increasing domain gaps: proteins that are within the same Family, Superfam, and Fold as the training set respectively. The results are in Table \ref{domain shift}. It is pertinent to note that MSA transformer's performance decreases dramatically when the gap between the domains increases. Our model surpasses MSA Transformer by a large margin on shifted domains, especially from 0.5032 to 0.6770 on Superfam. Our model proves to be more reliable for domain shifts, illustrating that retrieval facilitates the transfer across domains. }

Furthermore, we test our model on 108 out-of-domain De Novo proteins for the contact prediction task. De Novo proteins are synthesized by humans and have a different distribution from natural proteins. It can be seen in Figure \ref{fig: de novo contact} that, in addition to surpassing MSA transformer on average precision by 1\%, RSA also exceeds MSA transformer on 63.8\% of data, demonstrating that RSA is more capable of locating augmentations for out-of-distribution proteins.
We also test our model on the secondary structure task with new domain data, as shown in Appendix (Table~\ref{casp12} and Figure~\ref{fig:denovo ssp}). The results also show that our model surpasses MSA Transformer in transferring to unseen domains.

\input{table/04_alignment.tex}

\subsection{Retrieval Speed}
A severe speed bottleneck limits the use of previous MSA-based methods. In this part, we compare the computation time of RSA with MSA and an accelerated version of MSA as introduced in Section ~\ref{accelerated msa}. As shown in Figure \ref{fig: speed compare}, alignment time cost is much more intense than retrieval time. Even after reducing the alignment database size to 500, accelerated MSA still need 270 min to build MSA. At the same time RSA only uses dense retrieval, and is accelerated 373 times. Note that with extensive search, MSA can find \textit{all} available alignments in a database. However, this would be less beneficial to deep protein language models as the memory limit only suffices a few dozens of retrieved sequences.

\subsection{Retrieved Protein Interpretability}
\label{Retrieved Protein Interpretability}

The previous retrieval-augmented language models rely on a dense retriever to retrieve knowledge-relevant documents. However, it remains indistinct what constitutes knowledge for protein understanding and how retrieved sequences can be used for improving protein representations. In this section, we take a close look at the retrieved protein sequences to examine their homology and geometric properties.

\begin{figure}[t]
    \centering
    
    \includegraphics[width=0.7\linewidth]{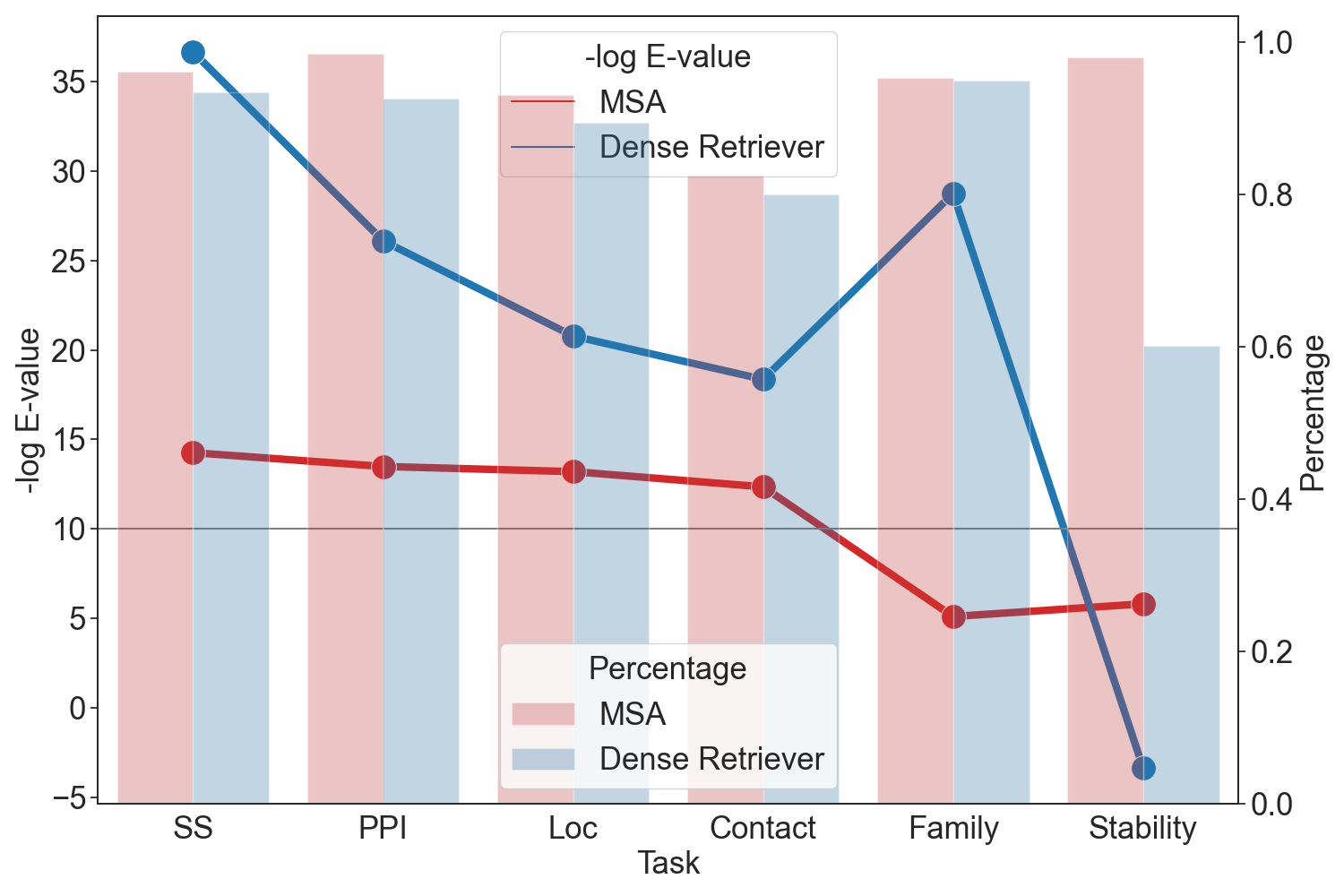}
        \caption{Plot of the -log(E-values) of MSA and Dense Retriever obtained sequences on the test sets for six tasks. E-values of both methods are obtained with HHblits\citep{remmert2012hhblits}. Sequences with -log E-value \textgreater 10 are high-quality homologous sequences. We also show with bar plots the percentage of sequences in the test sets that have homologous sequences.}
        \label{retrieval plot}
        \vspace{-5mm}
\end{figure}

\textbf{Dense Retrievers Find Homologous Sequences.} One type of knowledge distinct to the protein domain is sequence homology, which infers knowledge on shared ancestry between proteins in evolution. Homologous sequences are more likely to share functions or similar structures. We analyze whether retrieved sequences are homologous.

\begin{figure}[t]
    \centering
    
    \includegraphics[width=0.9\linewidth]{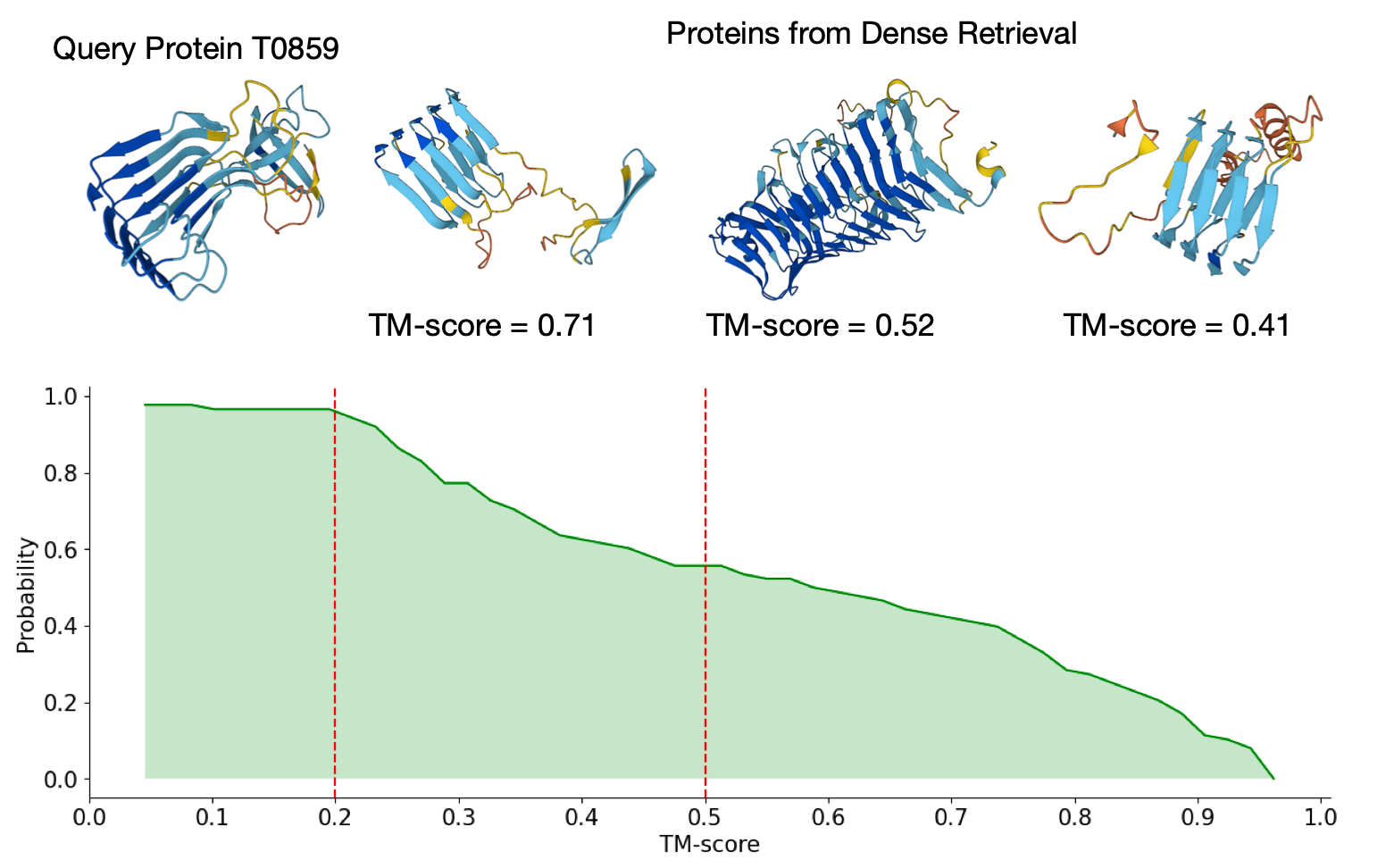}
        \caption{Plot of the cumulative distribution of TM-scores for proteins from dense retrieval. The value at $a$  shows the probability that TM-score is larger than $a$. We also give a visual example of retrieved protein to illustrate  similar structures.}
        \label{tm plot}
    \vspace{-3mm}
\end{figure}

{As illustrated in Figure \ref{retrieval plot} (right axis), across all six datasets, our dense retriever retrieved a high percentage of homologous proteins that can be aligned to the original protein sequence, comparable to traditional HMM-based MSA retrievers. We additionally plot each dataset's negative log E-values distribution in Figure \ref {retrieval plot}. Accordingly, pre-trained protein models can be used directly as dense retrieval of homologous sequences.} 
\input{table/05_variants.tex}
\textbf{RSA Retriever Find Structurally Similar Protein}
Protein structures are also central to protein functions and properties. In this section, we analyze whether retrieved sequences are structurally similar. In Figure \ref{tm plot}, we plot the TM scores between the RSA retrieved protein and the origin protein on ProteinNet \citep{alquraishi2019proteinnet} test set. Using ESMFold\footnote{https://esmatlas.com/resources?action=fold}, we obtain the 3D structures of the top 5 retrieved proteins and then calculate the TM score between these proteins and the query protein. Most of the retrieved proteins exceed the 0.2 criteria, which indicates structural similarity, and about half are above the 0.5 criteria, which indicates high quality. Accordingly, this indicates that the dense retrieval algorithm is capable of finding proteins with structural knowledge.

\subsection{Ablation Study}

\label{Ablation Study}
\textbf{Ablation on Retriever: Unaligned MSA Augmentation. }
We ablate RSA retriever by using MSA retrieved proteins as augmentations to our model, denoted as Unaligned MSA Augmentation. The results are in Table \ref{Unaligned MSA Augmentation}. As the result shows, Unaligned MSA Augmentation performs worse than our RSA model, especially on the Stability dataset, where the performance drops from 0.778 to 0.7443. It thus confirms the ability of our dense retriever to provide more abundant knowledge for protein models.

{\textbf{Ablation on Retriever: Ablation on Retrieval Number }
Our study examines the effect of injected knowledge quantity for RSA and all retrieval baselines. The results are listed in Table \ref{augmentation number}. We select the Contact dataset because all baseline models are implemented on this dataset. RSA and all baselines perform consistently better as the retrieval number increases. Also, our model outperforms all baseline models for all augmentation numbers.}

\input{table/06_depth.tex}

\textbf{Ablation on aggregation: }We compare RSA with Accelerated MSA Transformer to evaluate whether our aggregation method is beneficial for learning protein representations. Note that only part of the retrieved sequences that satisfy homologous sequence criteria are selected and utilized during alignment.  As shown in Table \ref{accelerated MSA}, the performance of the Accelerated MSA Transformer drops a lot compared to RSA. In contrast to MSA type aggregation, which is restricted by token alignment, our aggregation is more flexible and can accommodate proteins with variant knowledge.

\textbf{Is MSA retriever necessary? }Table \ref{accelerated MSA} illustrates that Accelerated MSA Transformer performs near to MSA Transformer (MSA N=16) for most datasets, except for Stability and PPI on which our retriever failed to find enough homologous sequences, as Figure \ref{retrieval plot} demonstrates. Our retriever is therefore capable of finding homologous sequences for most tasks and is able to replace the MSA retriever.

\textbf{Is MSA alignment necessary? }\label{section: msa alignment experiments}To support that MSA alignment is not necessary, we compare Unaligned MSA Augmentation to the original MSA transformer. As revealed by the results in Table \ref{Unaligned MSA Augmentation}. Unaligned MSA Augmentation performs close to the MSA transformer. This confirms our declaration that self-attention is capable of integrating protein sequences into representations.





%% file: table/08_domain_adpation.tex
\begin{table}[htbp]
\caption{The table shows remote homology prediction performance with increasing domain gaps: Family, Superfamily and Fold. }
\vspace{2mm}
\resizebox{\linewidth}{!}{
\setlength{\tabcolsep}{3.0mm}{
\begin{tabular}{lccc}
\toprule
Method  & Family & Superfam & Fold  \\ \midrule
Transformer                           &    0.101      & 0.518 &     0.078       \\
MSA Transformer (no MSA)         &      0.880     & 0.278 &    0.206        \\
ProtBERT         &      0.528      &   0.192 &    0.170       \\ \midrule

MSA Transformer         &      0.958       &   0.503 &  0.235       \\
Accelerated MSA Transformer &      0.945      &  0.406 &  0.227          \\
RSA (ProtBERT backbone)      &      \textbf{0.987}       &   \textbf{0.677}  & \textbf{0.267}        \\ 
\bottomrule
\end{tabular}}}
\vspace{-5mm}
\label{domain shift}
\end{table}

%% file: table/04_alignment.tex
\begin{table}[htbp]
\caption{Results for MSA Transformer and Unaligned MSA Augmentation on Homology and Stability task. Both models use MSA as inputs, but Unaligned MSA Augmentation unaligns MSA and augments the model by concatenating MSA sequence to the input.}
\vspace{2mm}
\resizebox{\linewidth}{!}{
\setlength{\tabcolsep}{3.0mm}{
\begin{tabular}{llll}
\toprule
Methods                    & Homology & Stability \\ \midrule
MSA Transformer            &  0.958    &  \textbf{0.796}\\
Unaligned MSA Augmentation &   0.973  & 0.749 \\ 
RSA                        &  \textbf{0.987}  &  0.778 \\
\bottomrule
\end{tabular}}}
\vspace{-2mm}
\label{Unaligned MSA Augmentation}
\end{table}

%% file: table/05_variants.tex

\begin{table}[htbp]
\vspace{-4mm}
\caption{Results for MSA Transformer and Accelerated MSA Transformer on downstream tasks. Accelerated MSA Transformer uses MSA built from dense retrieval sequences.}
\vspace{2mm}
\resizebox{\linewidth}{!}{
\setlength{\tabcolsep}{5.0mm}{
\begin{tabular}{cccc}
\toprule
Methods                    & MSA  & Accelerated MSA  & RSA \\
&Transformer & Transformer & \\
\midrule
SSP           &  0.654    &  0.634 & \textbf{0.691} \\
Contact       &   0.618  & 0.608 & \textbf{0.717} \\  
Homology       &   0.958  & 0.945 & \textbf{0.987} \\  
Stability       &   \textbf{0.796}  & 0.767 & 0.778\\  
Loc       &   0.694 & 0.682 & \textbf{0.795} \\  
PPI       &   0.751  & 0.679 & \textbf{0.827}\\  
\bottomrule
\end{tabular}
}
}

\label{accelerated MSA}
\end{table}

%% file: table/06_depth.tex
\begin{table}[htbp]
\vspace{-2mm}
\caption{The performance of retrieval augmentation models w.r.t. the number of retrieved sequences on contact prediction. }
\vspace{2mm}
\centering
\resizebox{\linewidth}{!}{
\setlength{\tabcolsep}{0.8mm}{
\begin{tabular}{lcccccc}
\toprule
Methods                     & N=1 & N=4 & N=8 & N=16 & N=32 & N= full \\ \midrule
Potts Model                 &  —   &  0.412   &  0.471   &  0.479    &   0.480   &    0.507     \\
MSA Transformer             & 0.397    &  0.579   &  0.560   &  0.618    &   0.669   & —       \\
Accelerated MSA Transformer &  0.397   &  0.524   &  0.538   &   0.608   &  0.654    &  —       \\
RSA                          & \textbf{0.556}    &  \textbf{0.595}   &  \textbf{0.615}   &  \textbf{0.717}        &  \textbf{0.719} &   —    \\
\bottomrule
\end{tabular}}}

\label{augmentation number}
\end{table}

%% file: 08_conclusions.tex
In this paper, we introduce a simple yet effective method to enhance protein representation learning. We demonstrate RSA as a fast yet high-performing method that has the potential to replace MSA-based methods in most scenarios. For future work, we hope to further scale up our RSA method and apply it to 3D folding tasks.

%% file: appendix.tex
\section{A Brief Recap on Proteins}
Proteins are the end products of the decoding process that starts with the information in cellular DNA. As workhorses of the cell, proteins compose structural and motor elements in the cell, and they serve as the catalysts for virtually every biochemical reaction that occurs in living things. This incredible array of functions derives from a startlingly simple code that specifies a hugely diverse set of structures.

In fact, each gene in cellular DNA contains the code for a unique protein structure. Not only are these proteins assembled with different amino acid sequences, but they also are held together by different bonds and folded into a variety of three-dimensional structures. The folded shape, or conformation, depends directly on the linear amino acid sequence of the protein.

\textbf{1. What are proteins made of?
} 

20 kinds of amino acids. Within a protein, multiple amino acids are linked together by peptide bonds, thereby forming a long chain.

\textbf{2. Protein structures
} 
There are four levels of structures:
\begin{itemize}
    \item Primary structure: amino acids sequence
    \item Secondary structure: stable folding patterns, including Alpha Helix, Beta Sheet.
    \item Tertiary structure: ensemble of formations and folds in a single linear chain of amino acids
    \item macromolecules with multiple polypeptide chains or subunits
\end{itemize}

\textbf{3. Protein Homology}
Protein homology is defined as shared ancestry in the evolutionary history of life. There exists different kinds of homology, including orthologous homology that may be similar function proteins across species (human and mice $\alpha$-goblin), and paralogous homology that is the result of mutations (human $\alpha$-goblin and $\beta$-goblin). Homologies result in conservative parts in protein sequences, or leads to similar structures and functions.

\textbf{4. Multiple Sequence Alignments}
A method used to determine conservative regions and find homologous sequences. An illustration is given here to show how sequences are aligned.

\section{Overview of Previous Protein Representation Augmentation Methods \label{augmentation methods}}

Below we introduce several state-of-the-art evolution augmentation methods for protein representation learning. These methods rely on MSA as input to extract representations. We use $x$ to denote a target protein and its MSA containing $N$ homologous proteins. 

\textbf{Potts Model~\citep{balakrishnan2011learning}}. This line of research fits a Markov Random Field to the underlying MSA with likelihood maximization. This approach is different from other protein representation learning methods as it only learns a pairwise score for residues contact prediction. We will focus on other methods that augment protein representations that can be used for diverse downstream predictions.

\textbf{Co-evolution Aggregator~\citep{yang2020improved, ju2021copulanet}}. One way to build an evolution informed representation is to use a MSA encoder to obtain the co-evolution related statistics. By applying MSA encoder on the $n$-th homologous protein in the MSA, we can get a total of $L\times d$ embeddings $R_n$, each position is a $d$ channel {one-hot} embedding indicating the amino acid type. We use $w_n$ to denote the weight from $R_n$ when computing the token representation $h_i$:
\begin{align}
    h_i = \frac{1}{M_{\textit{eff}}} \sum_{n=1}^N w_n R_n(i),
\end{align}
where $M_{\textit{eff}} =  \sum_{n=1}^N w_n$ and $w_n = \frac{1}{N}$. For contact prediction, pair co-evolution representation are computed in a similar way from the hadamard product: 
\begin{align}
    h_{ij} = \frac{1}{M_{\textit{eff}}} \sum_{n=1}^N w_n R_n(i) \bigotimes R_n(j).
\end{align}

\textbf{Ensembling Over MSA~\citep{rao2020transformer}}. This approach aligns and ensembles representations of homologous sequences. Consider the encoder extract the same token representations for unaligned and aligned sequences.
The ensembled token representation is:
\begin{align}
    h_i = \frac{1}{N}\sum_{n=1}^N R_n(i), h_{ij}= \frac{1}{N} \sum_{n=1}^N \sigma(\frac{R_n(i)W_Q(R_n(j)W_K)^T}{N\sqrt{d}}).
\end{align}

\textbf{MSA Transformer~\citep{rao2021msa}} In each transformer layer, a tied row attention encoder extracts the dense representation $R_n$, then a column attention encoder 
\begin{align}
   \label{eq: msa transformer}
   R_s(i) = \sum_{n=1}^N \sigma(\frac{R_s(i)W_Q(R_n(i)W_K)^T}{N\sqrt{d}})R_n(i)W_V.
\end{align}


\section{Experiment Setups}
\subsection {Introduction to the datasets}
\label{Introduction to the datasets}
\textit{Secondary structure prediction (SSP, 8-class)} aims to predict the secondary structure of proteins, which indicates the local structures. \textit{Contact prediction} predicts the long-range (distance \textgreater 6) residue-residue contact, which measures the ability of models to capture global tertiary structures. Homology  prediction aims to predict the fold label of any given protein, which indicates the evolutionary relationship of proteins. \textit{Stability} prediction is a protein engineering task, which measures the change in stability w.r.t. 
residue mutations. \textit{Subcellular Localization (Loc)} prediction predicts the local environment of proteins in the cell, which is closely related to protein functions and roles in biological processes. \textit{Protein protein interaction (PPI)} predicts whether two proteins interact with each other, which is crucial for protein function understanding and drug discovery. 

\begin{figure}[t]
    \centering
    
    \includegraphics[width=0.5\linewidth]{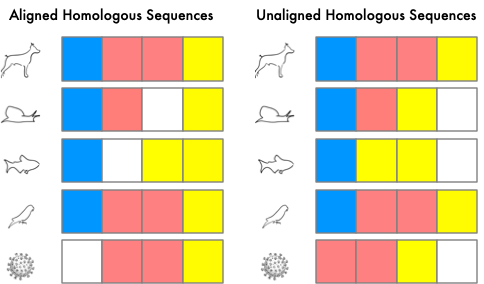}
        \caption{Illustrated difference of aligned and unaligned homologous sequences. }
        \label{fig: illustrated msa}
\end{figure}

\subsection{Retriever and MSA Details \label{Introduction to the retrievers}}
We adopt Faiss~\citep{johnson2019billion} indexing to accelerate the retrieval process by clustering the pre-trained dense vectors. In our implementation, we use the Inverted file with Product Quantizer encoding Indexing and set the size of quantized vectors to 64, the number of centroids to 4096, and the number of probes to 8. During retrieval, L2 distances are used to measure sequence similarity. The index is first trained on $.5\%$ of all retrieval data and then add all vectors. For MSA datasets, We use HHblits~\citep{remmert2012hhblits} to perform alignment, and the iteration and E-value thresholds of HHblits are set as $3$ and $1$. 

\section{Supplementary Experiment Analysis}

\subsection{Baselines}
\textbf{Protein representation learning benefits from knowledge augmentations.} 
In this part, we examine the performance of three types of baseline models. As shown in Table \ref{main result}, structure and evolution-related tasks all benefit greatly from pre-training, with over 20\% improvement in contact prediction and over 40\% improvement in homology prediction. Also, we observe that all kinds of knowledge-augmentation methods improve performance on a few downstream tasks. Though based purely on MSA information, Potts model shows competitive performance to vanilla pre-trained models. MSA Transformer with depth=16 MSA input also sees 12\% improvement on its no-MSA input performance. OntoProtein also improves on homology prediction and stability prediction, since knowledge graph enhancement is more suitable to function prediction than structure understanding. PMLM is the SOTA model on both structure and evolution-related tasks through co-evolution pre-training on Pfam database. This trend shows that current scale ( \textless 1 Billion parameters) pre-trained models still need knowledge augmentations to reach SOTA, and evolutionary knowledge is especially important for downstream prediction. 

\subsection{Domain Adaptation Analysis}

\begin{figure}
    \centering
    \resizebox{\linewidth}{!}{\includegraphics{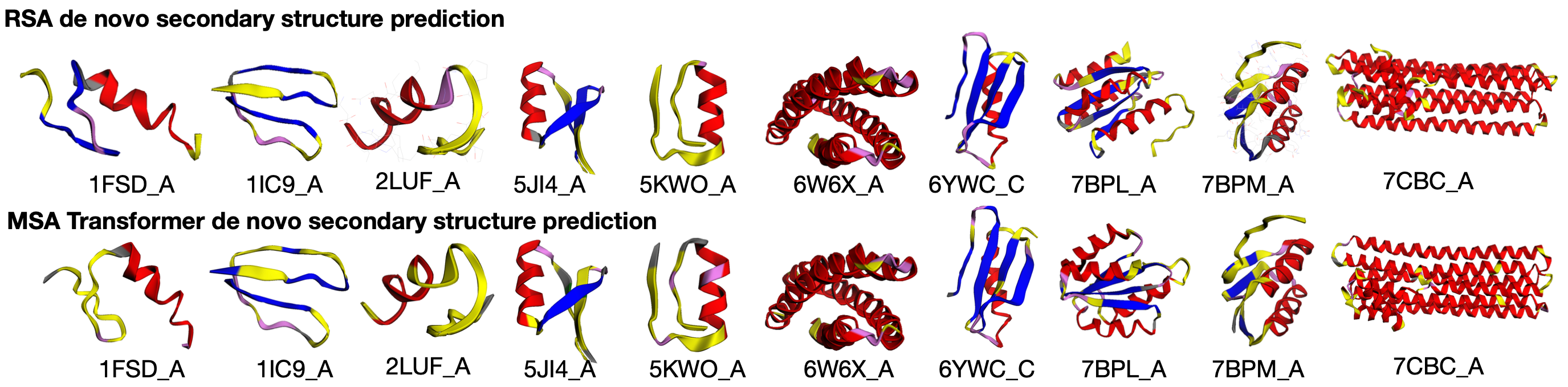}}
    \caption{Prediction of Secondary Structure on De Novo Dataset. Each color corresponds to a different secondary structure.}
    \label{fig:denovo ssp}
\end{figure}

In this section, we perform additional analysis on secondary structure prediction tasks. We perform training on NetSurfP-2.0\citep{klausen2019netsurfp} training set and test on two datasets with domain gaps. On CASP12, RSA marginally outperforms other baselines, as shown in Table 8. We also test on 10 de novo proteins (6YWC, 2LUF, 7BPM,  7BPL, 7CBC, 1FSD, 1IC9, 5JI4, 5KWO, 6W6X). Since we didn't find secondary structure labels for these proteins, we provide visualization in Figure~\ref{fig:denovo ssp} which shows that our model has an obvious overhead over MSA Transformer on predicting geometric components.

\begin{table}[htbp]
\caption{The domain adaptation performance of models on CASP12 secondary structure prediction.}
\vspace{2mm}
\centering
\begin{tabular}{lccc}
\toprule
Method   & CASP12   \\ \midrule
ProtBERT    &   0.628     \\
MSA Transformer  &    0.621        \\
Accelerated MSA Transformer  &     0.620         \\
RSA (ProtBERT backbone)      &  \textbf{0.631}          \\ 
\bottomrule
\end{tabular}

\label{casp12}
\end{table}

\section{Dataset details}
\subsection{Downstream tasks}
Table~\ref{Table: table task} gives the details for the datasets. 
\input{table/02_tasks.tex}

\subsection{De Novo Protein Dataset}
We follow \citet{chowdhury2022single} to curate a de novo dataset of 108 proteins from Protein Data Bank~\citep{bankrcsb}. These proteins are originally designed de novo using computationally parametrized energy functions and are  well-suited for out-of-domain tests. Note that different from orphan dataset, MSA can be built for this dataset, though showing a decline in quality.

\section{Additional Visualization of Retrieved Sequence 3D Structure}

\begin{figure}[htbp]
    \centering
    \resizebox{0.5\linewidth}{!}{\includegraphics{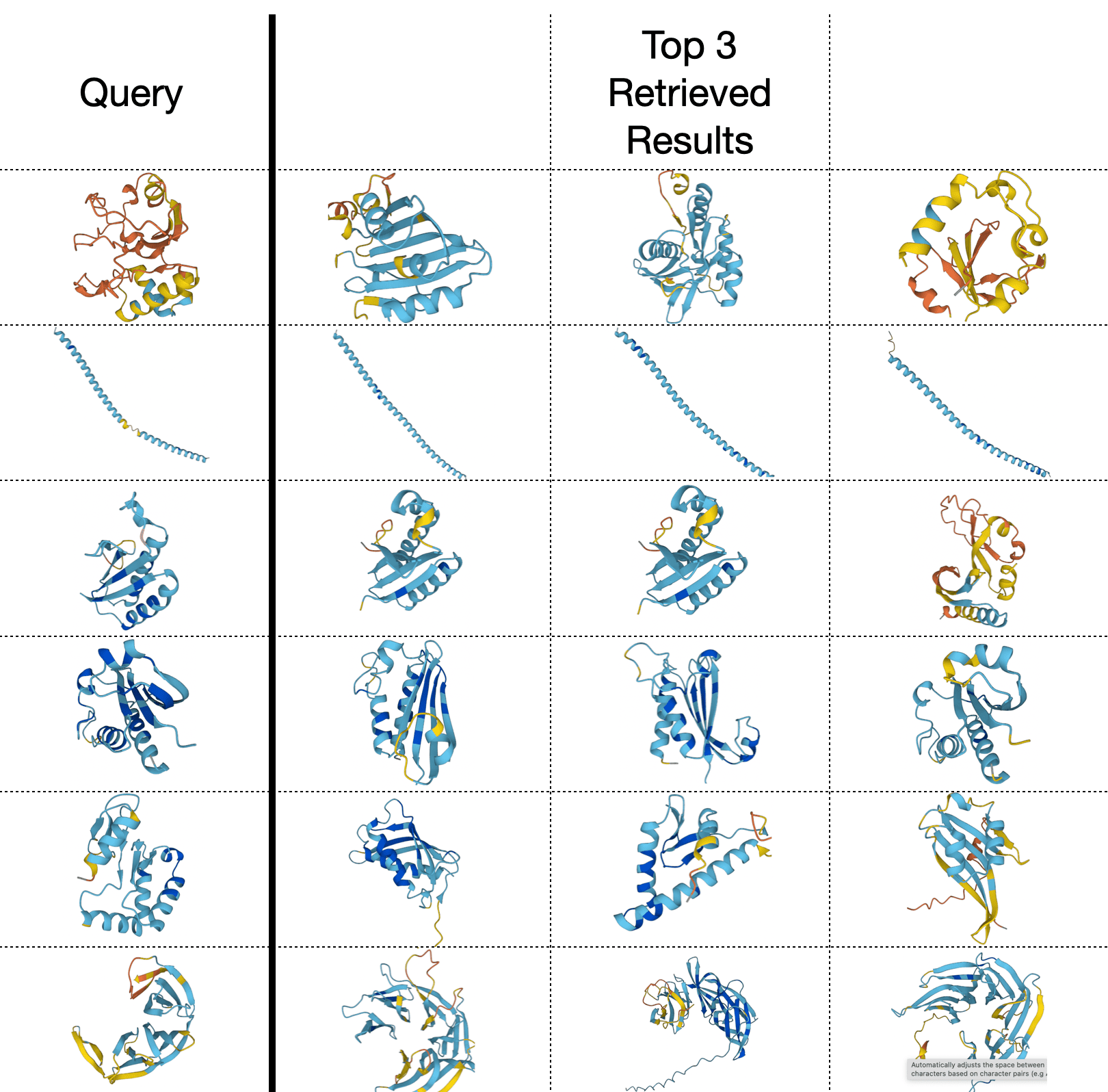}}
    \caption{Query and Retrieved Sequence Structures}
    \label{fig:structures}
\end{figure}

As shown in Figure~\ref{fig:structures}, we random picked a few more examples to illustrate the structural similarity between query protein and retrieval proteins.

%% file: table/02_tasks.tex
\begin{table*}[ht]
\centering
\caption{Overview for datasets in downstream tasks}
\vspace{2mm}
\small
\begin{tabular}{llll}
\hline
Task Name                      & Dataset source                                           & \#train sequences & \#test sequences \\ \hline
Secondary Structure Prediction & NetSurfP-2.0~\citep{klausen2019netsurfp}                                            & 8,678             & 513              \\ 
Contact Prediction             & ProteinNet~\citep{alquraishi2019proteinnet}                                                & 25,299            & 40               \\ 
Remote Homology Prediction     & Deepsf~\citep{hou2018deepsf}                                                   & 12,312            & 718              \\ 
Stability Prediction           & Rocklin's Dataset~\citep{rocklin2017global}& 53,571            & 12,851           \\ 
Subcellular Localization       & DeepLoc~\citep{almagro2017deeploc}                                                  & 8,945             & 2,768            \\ 
Protein Protein Interaction    & Pan's Dataset~\citep{pan2010large}                                              & 6,844             & 227              \\ \hline
\end{tabular}

\label{Table: table task}
\end{table*}

%% file: main.bbl
\begin{thebibliography}{63}
\providecommand{\natexlab}[1]{#1}
\providecommand{\url}[1]{\texttt{#1}}
\expandafter\ifx\csname urlstyle\endcsname\relax
  \providecommand{\doi}[1]{doi: #1}\else
  \providecommand{\doi}{doi: \begingroup \urlstyle{rm}\Url}\fi

\bibitem[Alley et~al.(2019)Alley, Khimulya, Biswas, AlQuraishi, and
  Church]{alley2019unified}
Alley, E.~C., Khimulya, G., Biswas, S., AlQuraishi, M., and Church, G.~M.
\newblock Unified rational protein engineering with sequence-based deep
  representation learning.
\newblock \emph{Nature methods}, 16\penalty0 (12):\penalty0 1315--1322, 2019.

\bibitem[Almagro~Armenteros et~al.(2017)Almagro~Armenteros, S{\o}nderby,
  S{\o}nderby, Nielsen, and Winther]{almagro2017deeploc}
Almagro~Armenteros, J.~J., S{\o}nderby, C.~K., S{\o}nderby, S.~K., Nielsen, H.,
  and Winther, O.
\newblock Deeploc: prediction of protein subcellular localization using deep
  learning.
\newblock \emph{Bioinformatics}, 33\penalty0 (21):\penalty0 3387--3395, 2017.

\bibitem[AlQuraishi(2019)]{alquraishi2019proteinnet}
AlQuraishi, M.
\newblock Proteinnet: a standardized data set for machine learning of protein
  structure.
\newblock \emph{BMC bioinformatics}, 20\penalty0 (1):\penalty0 1--10, 2019.

\bibitem[Altschul \& Koonin(1998)Altschul and Koonin]{altschul1998iterated}
Altschul, S.~F. and Koonin, E.~V.
\newblock Iterated profile searches with psi-blast—a tool for discovery in
  protein databases.
\newblock \emph{Trends in biochemical sciences}, 23\penalty0 (11):\penalty0
  444--447, 1998.

\bibitem[Anishchenko et~al.(2021)Anishchenko, Pellock, Chidyausiku, Ramelot,
  Ovchinnikov, Hao, Bafna, Norn, Kang, Bera, et~al.]{anishchenko2021novo}
Anishchenko, I., Pellock, S.~J., Chidyausiku, T.~M., Ramelot, T.~A.,
  Ovchinnikov, S., Hao, J., Bafna, K., Norn, C., Kang, A., Bera, A.~K., et~al.
\newblock De novo protein design by deep network hallucination.
\newblock \emph{Nature}, 600\penalty0 (7889):\penalty0 547--552, 2021.

\bibitem[Apweiler et~al.(2004)Apweiler, Bairoch, Wu, Barker, Boeckmann, Ferro,
  Gasteiger, Huang, Lopez, Magrane, et~al.]{apweiler2004uniprot}
Apweiler, R., Bairoch, A., Wu, C.~H., Barker, W.~C., Boeckmann, B., Ferro, S.,
  Gasteiger, E., Huang, H., Lopez, R., Magrane, M., et~al.
\newblock Uniprot: the universal protein knowledgebase.
\newblock \emph{Nucleic acids research}, 32\penalty0 (suppl\_1):\penalty0
  D115--D119, 2004.

\bibitem[Balakrishnan et~al.(2011)Balakrishnan, Kamisetty, Carbonell, Lee, and
  Langmead]{balakrishnan2011learning}
Balakrishnan, S., Kamisetty, H., Carbonell, J.~G., Lee, S.-I., and Langmead,
  C.~J.
\newblock Learning generative models for protein fold families.
\newblock \emph{Proteins: Structure, Function, and Bioinformatics}, 79\penalty0
  (4):\penalty0 1061--1078, 2011.

\bibitem[Bank(2022)]{bankrcsb}
Bank, P.~D.
\newblock Rcsb pdb. 2022, 2022.

\bibitem[Borgeaud et~al.(2022)Borgeaud, Mensch, Hoffmann, Cai, Rutherford,
  Millican, Van Den~Driessche, Lespiau, Damoc, Clark,
  et~al.]{borgeaud2022improving}
Borgeaud, S., Mensch, A., Hoffmann, J., Cai, T., Rutherford, E., Millican, K.,
  Van Den~Driessche, G.~B., Lespiau, J.-B., Damoc, B., Clark, A., et~al.
\newblock Improving language models by retrieving from trillions of tokens.
\newblock In \emph{International conference on machine learning}, pp.\
  2206--2240. PMLR, 2022.

\bibitem[Chowdhury et~al.(2022)Chowdhury, Bouatta, Biswas, Floristean, Kharkar,
  Roy, Rochereau, Ahdritz, Zhang, Church, et~al.]{chowdhury2022single}
Chowdhury, R., Bouatta, N., Biswas, S., Floristean, C., Kharkar, A., Roy, K.,
  Rochereau, C., Ahdritz, G., Zhang, J., Church, G.~M., et~al.
\newblock Single-sequence protein structure prediction using a language model
  and deep learning.
\newblock \emph{Nature Biotechnology}, 40\penalty0 (11):\penalty0 1617--1623,
  2022.

\bibitem[Deorowicz et~al.(2016)Deorowicz, Debudaj-Grabysz, and
  Gudy{\'s}]{deorowicz2016famsa}
Deorowicz, S., Debudaj-Grabysz, A., and Gudy{\'s}, A.
\newblock Famsa: Fast and accurate multiple sequence alignment of huge protein
  families.
\newblock \emph{Scientific reports}, 6\penalty0 (1):\penalty0 1--13, 2016.

\bibitem[El-Gebali et~al.(2018)El-Gebali, Mistry, Bateman, Eddy, Luciani,
  Potter, Qureshi, Richardson, Salazar, Smart, Sonnhammer, Hirsh, Paladin,
  Piovesan, Tosatto, and Finn]{pfam-gebali2018}
El-Gebali, S., Mistry, J., Bateman, A., Eddy, S.~R., Luciani, A., Potter,
  S.~C., Qureshi, M., Richardson, L.~J., Salazar, G.~A., Smart, A., Sonnhammer,
  E.~L., Hirsh, L., Paladin, L., Piovesan, D., Tosatto, S.~C., and Finn, R.~D.
\newblock {The Pfam protein families database in 2019}.
\newblock \emph{Nucleic Acids Research}, 47\penalty0 (D1):\penalty0 D427--D432,
  10 2018.
\newblock ISSN 0305-1048.
\newblock \doi{10.1093/nar/gky995}.
\newblock URL \url{https://doi.org/10.1093/nar/gky995}.

\bibitem[Elnaggar et~al.(2020)Elnaggar, Heinzinger, Dallago, Rihawi, Wang,
  Jones, Gibbs, Feher, Angerer, Steinegger, et~al.]{elnaggar2020prottrans}
Elnaggar, A., Heinzinger, M., Dallago, C., Rihawi, G., Wang, Y., Jones, L.,
  Gibbs, T., Feher, T., Angerer, C., Steinegger, M., et~al.
\newblock Prottrans: towards cracking the language of life's code through
  self-supervised deep learning and high performance computing.
\newblock \emph{arXiv preprint arXiv:2007.06225}, 2020.

\bibitem[Elnaggar et~al.(2021)Elnaggar, Heinzinger, Dallago, Rehawi, Wang,
  Jones, Gibbs, Feher, Angerer, Steinegger, Bhowmik, and
  Rost]{AhmedElnaggar2021ProtTransTC}
Elnaggar, A., Heinzinger, M., Dallago, C., Rehawi, G., Wang, Y., Jones, L.,
  Gibbs, T., Feher, T., Angerer, C., Steinegger, M., Bhowmik, D., and Rost, B.
\newblock Prottrans: Towards cracking the language of life’s code through
  self-supervised learning.
\newblock \emph{bioRxiv}, 2021.

\bibitem[Fang et~al.(2022)Fang, Wang, Liu, He, Lin, Xiang, Zhang, Wu, Li, and
  Song]{fang2022helixfold}
Fang, X., Wang, F., Liu, L., He, J., Lin, D., Xiang, Y., Zhang, X., Wu, H., Li,
  H., and Song, L.
\newblock Helixfold-single: Msa-free protein structure prediction by using
  protein language model as an alternative.
\newblock \emph{arXiv preprint arXiv:2207.13921}, 2022.

\bibitem[Garrett \& Grisham(2016)Garrett and Grisham]{garrett2016biochemistry}
Garrett, R.~H. and Grisham, C.~M.
\newblock \emph{Biochemistry}.
\newblock Cengage Learning, 2016.

\bibitem[Goyal et~al.(2022)Goyal, Friesen, Banino, Weber, Ke, Badia, Guez,
  Mirza, Humphreys, Konyushova, et~al.]{AnirudhGoyal2022RetrievalAugmentedRL}
Goyal, A., Friesen, A., Banino, A., Weber, T., Ke, N.~R., Badia, A.~P., Guez,
  A., Mirza, M., Humphreys, P.~C., Konyushova, K., et~al.
\newblock Retrieval-augmented reinforcement learning.
\newblock In \emph{International Conference on Machine Learning}, pp.\
  7740--7765. PMLR, 2022.

\bibitem[Guu et~al.(2020{\natexlab{a}})Guu, Lee, Tung, Pasupat, and
  Chang]{guu2020retrieval}
Guu, K., Lee, K., Tung, Z., Pasupat, P., and Chang, M.
\newblock Retrieval augmented language model pre-training.
\newblock In \emph{International Conference on Machine Learning}, pp.\
  3929--3938. PMLR, 2020{\natexlab{a}}.

\bibitem[Guu et~al.(2020{\natexlab{b}})Guu, Lee, Tung, Pasupat, and
  Chang]{KelvinGuu2020REALMRL}
Guu, K., Lee, K., Tung, Z., Pasupat, P., and Chang, M.-W.
\newblock Realm: Retrieval-augmented language model pre-training.
\newblock \emph{international conference on machine learning},
  2020{\natexlab{b}}.

\bibitem[Hauser et~al.(2016)Hauser, Steinegger, and
  S{\"o}ding]{hauser2016mmseqs}
Hauser, M., Steinegger, M., and S{\"o}ding, J.
\newblock Mmseqs software suite for fast and deep clustering and searching of
  large protein sequence sets.
\newblock \emph{Bioinformatics}, 32\penalty0 (9):\penalty0 1323--1330, 2016.

\bibitem[He et~al.(2021{\natexlab{a}})He, Neubig, and
  Berg-Kirkpatrick]{he2021efficient}
He, J., Neubig, G., and Berg-Kirkpatrick, T.
\newblock Efficient nearest neighbor language models.
\newblock \emph{arXiv preprint arXiv:2109.04212}, 2021{\natexlab{a}}.

\bibitem[He et~al.(2021{\natexlab{b}})He, Zhang, Wu, Xia, Ju, Zhang, Liu, Xia,
  Zhu, Deng, et~al.]{he2021pre}
He, L., Zhang, S., Wu, L., Xia, H., Ju, F., Zhang, H., Liu, S., Xia, Y., Zhu,
  J., Deng, P., et~al.
\newblock Pre-training co-evolutionary protein representation via a pairwise
  masked language model.
\newblock \emph{arXiv preprint arXiv:2110.15527}, 2021{\natexlab{b}}.

\bibitem[Heinzinger et~al.(2019)Heinzinger, Elnaggar, Wang, Dallago, Nechaev,
  Matthes, and Rost]{heinzinger2019modeling}
Heinzinger, M., Elnaggar, A., Wang, Y., Dallago, C., Nechaev, D., Matthes, F.,
  and Rost, B.
\newblock Modeling aspects of the language of life through transfer-learning
  protein sequences.
\newblock \emph{BMC bioinformatics}, 20\penalty0 (1):\penalty0 1--17, 2019.

\bibitem[Hong et~al.(2021)Hong, Sun, Zheng, Tan, and Li]{hong2021fastmsa}
Hong, L., Sun, S., Zheng, L., Tan, Q., and Li, Y.
\newblock fastmsa: Accelerating multiple sequence alignment with dense
  retrieval on protein language.
\newblock \emph{bioRxiv}, 2021.

\bibitem[Hong et~al.(2022)Hong, Song, Ko, Lee, and Shin]{hong2022s}
Hong, Y., Song, J., Ko, J., Lee, J., and Shin, W.-H.
\newblock S-pred: protein structural property prediction using msa transformer.
\newblock \emph{Scientific reports}, 12\penalty0 (1):\penalty0 1--11, 2022.

\bibitem[Hou et~al.(2018)Hou, Adhikari, and Cheng]{hou2018deepsf}
Hou, J., Adhikari, B., and Cheng, J.
\newblock Deepsf: deep convolutional neural network for mapping protein
  sequences to folds.
\newblock \emph{Bioinformatics}, 34\penalty0 (8):\penalty0 1295--1303, 2018.

\bibitem[Hu et~al.(2022)Hu, Yuan, Yang, Ju, Su, Wang, Yang, and
  Ding]{hu2022exploring}
Hu, M., Yuan, F., Yang, K.~K., Ju, F., Su, J., Wang, H., Yang, F., and Ding, Q.
\newblock Exploring evolution-aware \& -free protein language models as protein
  function predictors.
\newblock In Oh, A.~H., Agarwal, A., Belgrave, D., and Cho, K. (eds.),
  \emph{Advances in Neural Information Processing Systems}, 2022.
\newblock URL \url{https://openreview.net/forum?id=U8k0QaBgXS}.

\bibitem[Johnson et~al.(2019{\natexlab{a}})Johnson, Douze, and
  J{\'e}gou]{faiss}
Johnson, J., Douze, M., and J{\'e}gou, H.
\newblock Billion-scale similarity search with {GPUs}.
\newblock \emph{IEEE Transactions on Big Data}, 7\penalty0 (3):\penalty0
  535--547, 2019{\natexlab{a}}.

\bibitem[Johnson et~al.(2019{\natexlab{b}})Johnson, Douze, and
  J{\'e}gou]{johnson2019billion}
Johnson, J., Douze, M., and J{\'e}gou, H.
\newblock Billion-scale similarity search with {GPUs}.
\newblock \emph{IEEE Transactions on Big Data}, 7\penalty0 (3):\penalty0
  535--547, 2019{\natexlab{b}}.

\bibitem[Johnson et~al.(2010)Johnson, Eddy, and Portugaly]{johnson2010hidden}
Johnson, L.~S., Eddy, S.~R., and Portugaly, E.
\newblock Hidden markov model speed heuristic and iterative hmm search
  procedure.
\newblock \emph{BMC bioinformatics}, 11\penalty0 (1):\penalty0 1--8, 2010.

\bibitem[Ju et~al.(2021)Ju, Zhu, Shao, Kong, Liu, Zheng, and
  Bu]{ju2021copulanet}
Ju, F., Zhu, J., Shao, B., Kong, L., Liu, T.-Y., Zheng, W.-M., and Bu, D.
\newblock Copulanet: Learning residue co-evolution directly from multiple
  sequence alignment for protein structure prediction.
\newblock \emph{Nature communications}, 12\penalty0 (1):\penalty0 1--9, 2021.

\bibitem[Jumper et~al.(2021)Jumper, Evans, Pritzel, Green, Figurnov,
  Ronneberger, Tunyasuvunakool, Bates, {\v{Z}}{\'\i}dek, Potapenko,
  et~al.]{jumper2021highly}
Jumper, J., Evans, R., Pritzel, A., Green, T., Figurnov, M., Ronneberger, O.,
  Tunyasuvunakool, K., Bates, R., {\v{Z}}{\'\i}dek, A., Potapenko, A., et~al.
\newblock Highly accurate protein structure prediction with alphafold.
\newblock \emph{Nature}, 596\penalty0 (7873):\penalty0 583--589, 2021.

\bibitem[Kamisetty et~al.(2013)Kamisetty, Ovchinnikov, and
  Baker]{HetunandanKamisetty2013AssessingTU}
Kamisetty, H., Ovchinnikov, S., and Baker, D.
\newblock Assessing the utility of coevolution-based residue-residue contact
  predictions in a sequence- and structure-rich era.
\newblock \emph{Proceedings of the National Academy of Sciences of the United
  States of America}, 2013.

\bibitem[Kaplan et~al.(2020)Kaplan, McCandlish, Henighan, Brown, Chess, Child,
  Gray, Radford, Wu, and Amodei]{kaplan2020scaling}
Kaplan, J., McCandlish, S., Henighan, T., Brown, T.~B., Chess, B., Child, R.,
  Gray, S., Radford, A., Wu, J., and Amodei, D.
\newblock Scaling laws for neural language models.
\newblock \emph{arXiv preprint arXiv:2001.08361}, 2020.

\bibitem[Khandelwal et~al.(2019)Khandelwal, Levy, Jurafsky, Zettlemoyer, and
  Lewis]{UrvashiKhandelwal2019GeneralizationTM}
Khandelwal, U., Levy, O., Jurafsky, D., Zettlemoyer, L., and Lewis, M.
\newblock Generalization through memorization: Nearest neighbor language
  models.
\newblock \emph{Learning}, 2019.

\bibitem[Klausen et~al.(2019)Klausen, Jespersen, Nielsen, Jensen, Jurtz,
  Soenderby, Sommer, Winther, Nielsen, Petersen, et~al.]{klausen2019netsurfp}
Klausen, M.~S., Jespersen, M.~C., Nielsen, H., Jensen, K.~K., Jurtz, V.~I.,
  Soenderby, C.~K., Sommer, M. O.~A., Winther, O., Nielsen, M., Petersen, B.,
  et~al.
\newblock Netsurfp-2.0: Improved prediction of protein structural features by
  integrated deep learning.
\newblock \emph{Proteins: Structure, Function, and Bioinformatics}, 87\penalty0
  (6):\penalty0 520--527, 2019.

\bibitem[Korendovych \& DeGrado(2020)Korendovych and
  DeGrado]{korendovych2020novo}
Korendovych, I.~V. and DeGrado, W.~F.
\newblock De novo protein design, a retrospective.
\newblock \emph{Quarterly reviews of biophysics}, 53, 2020.

\bibitem[Lin et~al.(2022)Lin, Akin, Rao, Hie, Zhu, Lu, dos Santos~Costa,
  Fazel-Zarandi, Sercu, Candido, et~al.]{lin2022language}
Lin, Z., Akin, H., Rao, R., Hie, B., Zhu, Z., Lu, W., dos Santos~Costa, A.,
  Fazel-Zarandi, M., Sercu, T., Candido, S., et~al.
\newblock Language models of protein sequences at the scale of evolution enable
  accurate structure prediction.
\newblock \emph{bioRxiv}, 2022.

\bibitem[Liu(2017)]{liu2017deep}
Liu, X.
\newblock Deep recurrent neural network for protein function prediction from
  sequence.
\newblock \emph{arXiv preprint arXiv:1701.08318}, 2017.

\bibitem[Marks et~al.(2011)Marks, Colwell, Sheridan, Hopf, Pagnani, Zecchina,
  and Sander]{marks2011protein}
Marks, D.~S., Colwell, L.~J., Sheridan, R., Hopf, T.~A., Pagnani, A., Zecchina,
  R., and Sander, C.
\newblock Protein 3d structure computed from evolutionary sequence variation.
\newblock \emph{PloS one}, 6\penalty0 (12):\penalty0 e28766, 2011.

\bibitem[Morcos et~al.(2011)Morcos, Pagnani, Lunt, Bertolino, Marks, Sander,
  Zecchina, Onuchic, Hwa, and Weigt]{morcos2011direct}
Morcos, F., Pagnani, A., Lunt, B., Bertolino, A., Marks, D.~S., Sander, C.,
  Zecchina, R., Onuchic, J.~N., Hwa, T., and Weigt, M.
\newblock Direct-coupling analysis of residue coevolution captures native
  contacts across many protein families.
\newblock \emph{Proceedings of the National Academy of Sciences}, 108\penalty0
  (49):\penalty0 E1293--E1301, 2011.

\bibitem[O'Sullivan et~al.(2004)O'Sullivan, Suhre, Abergel, Higgins, and
  Notredame]{o20043dcoffee}
O'Sullivan, O., Suhre, K., Abergel, C., Higgins, D.~G., and Notredame, C.
\newblock 3dcoffee: combining protein sequences and structures within multiple
  sequence alignments.
\newblock \emph{Journal of molecular biology}, 340\penalty0 (2):\penalty0
  385--395, 2004.

\bibitem[Pan et~al.(2010)Pan, Zhang, and Shen]{pan2010large}
Pan, X.-Y., Zhang, Y.-N., and Shen, H.-B.
\newblock Large-scale prediction of human protein- protein interactions from
  amino acid sequence based on latent topic features.
\newblock \emph{Journal of proteome research}, 9\penalty0 (10):\penalty0
  4992--5001, 2010.

\bibitem[Perdig{\~a}o et~al.(2015)Perdig{\~a}o, Heinrich, Stolte, Sabir,
  Buckley, Tabor, Signal, Gloss, Hammang, Rost, et~al.]{perdigao2015unexpected}
Perdig{\~a}o, N., Heinrich, J., Stolte, C., Sabir, K.~S., Buckley, M.~J.,
  Tabor, B., Signal, B., Gloss, B.~S., Hammang, C.~J., Rost, B., et~al.
\newblock Unexpected features of the dark proteome.
\newblock \emph{Proceedings of the National Academy of Sciences}, 112\penalty0
  (52):\penalty0 15898--15903, 2015.

\bibitem[Rao et~al.(2019)Rao, Bhattacharya, Thomas, Duan, Chen, Canny, Abbeel,
  and Song]{rao2019evaluating}
Rao, R., Bhattacharya, N., Thomas, N., Duan, Y., Chen, P., Canny, J., Abbeel,
  P., and Song, Y.
\newblock Evaluating protein transfer learning with tape.
\newblock \emph{Advances in neural information processing systems}, 32, 2019.

\bibitem[Rao et~al.(2020)Rao, Meier, Sercu, Ovchinnikov, and
  Rives]{rao2020transformer}
Rao, R., Meier, J., Sercu, T., Ovchinnikov, S., and Rives, A.
\newblock Transformer protein language models are unsupervised structure
  learners.
\newblock \emph{Biorxiv}, 2020.

\bibitem[Rao et~al.(2021)Rao, Liu, Verkuil, Meier, Canny, Abbeel, Sercu, and
  Rives]{rao2021msa}
Rao, R.~M., Liu, J., Verkuil, R., Meier, J., Canny, J., Abbeel, P., Sercu, T.,
  and Rives, A.
\newblock Msa transformer.
\newblock In \emph{International Conference on Machine Learning}, pp.\
  8844--8856. PMLR, 2021.

\bibitem[Remmert et~al.(2012)Remmert, Biegert, Hauser, and
  S{\"o}ding]{remmert2012hhblits}
Remmert, M., Biegert, A., Hauser, A., and S{\"o}ding, J.
\newblock Hhblits: lightning-fast iterative protein sequence searching by
  hmm-hmm alignment.
\newblock \emph{Nature methods}, 9\penalty0 (2):\penalty0 173--175, 2012.

\bibitem[Riesselman et~al.(2019)Riesselman, Shin, Kollasch, McMahon, Simon,
  Sander, Manglik, Kruse, and Marks]{riesselman2019accelerating}
Riesselman, A., Shin, J.-E., Kollasch, A., McMahon, C., Simon, E., Sander, C.,
  Manglik, A., Kruse, A., and Marks, D.
\newblock Accelerating protein design using autoregressive generative models.
\newblock \emph{BioRxiv}, 757252, 2019.

\bibitem[Rives et~al.(2019)Rives, Goyal, Meier, Guo, Ott, Zitnick, Ma, and
  Fergus]{esm1b}
Rives, A., Goyal, S., Meier, J., Guo, D., Ott, M., Zitnick, C.~L., Ma, J., and
  Fergus, R.
\newblock Biological structure and function emerge from scaling unsupervised
  learning to 250 million protein sequences.
\newblock \emph{Proceedings of the National Academy of Sciences of the United
  States of America}, 2019.

\bibitem[Rocklin et~al.(2017)Rocklin, Chidyausiku, Goreshnik, Ford, Houliston,
  Lemak, Carter, Ravichandran, Mulligan, Chevalier, et~al.]{rocklin2017global}
Rocklin, G.~J., Chidyausiku, T.~M., Goreshnik, I., Ford, A., Houliston, S.,
  Lemak, A., Carter, L., Ravichandran, R., Mulligan, V.~K., Chevalier, A.,
  et~al.
\newblock Global analysis of protein folding using massively parallel design,
  synthesis, and testing.
\newblock \emph{Science}, 357\penalty0 (6347):\penalty0 168--175, 2017.

\bibitem[Roy et~al.(2010)Roy, Kucukural, and Zhang]{roy2010tasser}
Roy, A., Kucukural, A., and Zhang, Y.
\newblock I-tasser: a unified platform for automated protein structure and
  function prediction.
\newblock \emph{Nature protocols}, 5\penalty0 (4):\penalty0 725--738, 2010.

\bibitem[Sadowski \& Jones(2009)Sadowski and Jones]{sadowski2009sequence}
Sadowski, M. and Jones, D.
\newblock The sequence--structure relationship and protein function prediction.
\newblock \emph{Current opinion in structural biology}, 19\penalty0
  (3):\penalty0 357--362, 2009.

\bibitem[Stefani(2004)]{stefani2004protein}
Stefani, M.
\newblock Protein misfolding and aggregation: new examples in medicine and
  biology of the dark side of the protein world.
\newblock \emph{Biochimica et biophysica acta (BBA)-Molecular basis of
  disease}, 1739\penalty0 (1):\penalty0 5--25, 2004.

\bibitem[Vaswani et~al.(2017)Vaswani, Shazeer, Parmar, Uszkoreit, Jones, Gomez,
  Kaiser, and Polosukhin]{vaswani2017attention}
Vaswani, A., Shazeer, N., Parmar, N., Uszkoreit, J., Jones, L., Gomez, A.~N.,
  Kaiser, {\L}., and Polosukhin, I.
\newblock Attention is all you need.
\newblock \emph{Advances in neural information processing systems}, 30, 2017.

\bibitem[Wang et~al.(2022)Wang, Liu, Wang, Song, Tang, Le, Grau, and
  Liu]{DingminWang2022AugmentingMP}
Wang, D., Liu, S., Wang, H., Song, L., Tang, J., Le, S., Grau, B.~C., and Liu,
  Q.
\newblock Augmenting message passing by retrieving similar graphs.
\newblock \emph{arXiv preprint arXiv:2206.00362}, 2022.

\bibitem[Wu et~al.(2022)Wu, Ding, Wang, Shen, Zhang, Luo, Su, Wu, Xie, Berger,
  et~al.]{wu2022high}
Wu, R., Ding, F., Wang, R., Shen, R., Zhang, X., Luo, S., Su, C., Wu, Z., Xie,
  Q., Berger, B., et~al.
\newblock High-resolution de novo structure prediction from primary sequence.
\newblock \emph{BioRxiv}, pp.\  2022--07, 2022.

\bibitem[Xia et~al.(2009)Xia, Zhang, Su, and Sun]{xia2009micalign}
Xia, X., Zhang, S., Su, Y., and Sun, Z.
\newblock Micalign: a sequence-to-structure alignment tool integrating multiple
  sources of information in conditional random fields.
\newblock \emph{Bioinformatics}, 25\penalty0 (11):\penalty0 1433--1434, 2009.

\bibitem[Xu et~al.(2022)Xu, Zhang, Lu, Zhu, Zhang, Ma, Liu, and
  Tang]{xu2022peer}
Xu, M., Zhang, Z., Lu, J., Zhu, Z., Zhang, Y., Ma, C., Liu, R., and Tang, J.
\newblock Peer: A comprehensive and multi-task benchmark for protein sequence
  understanding.
\newblock \emph{arXiv preprint arXiv:2206.02096}, 2022.

\bibitem[Yang et~al.(2020)Yang, Anishchenko, Park, Peng, Ovchinnikov, and
  Baker]{yang2020improved}
Yang, J., Anishchenko, I., Park, H., Peng, Z., Ovchinnikov, S., and Baker, D.
\newblock Improved protein structure prediction using predicted interresidue
  orientations.
\newblock \emph{Proceedings of the National Academy of Sciences}, 117\penalty0
  (3):\penalty0 1496--1503, 2020.

\bibitem[Ye et~al.(2006)Ye, McGinnis, and Madden]{ye2006blast}
Ye, J., McGinnis, S., and Madden, T.~L.
\newblock Blast: improvements for better sequence analysis.
\newblock \emph{Nucleic acids research}, 34\penalty0 (suppl\_2):\penalty0
  W6--W9, 2006.

\bibitem[Zhang et~al.(2021)Zhang, Ju, Zhu, He, Shao, Zheng, and
  Liu]{zhang2021co}
Zhang, H., Ju, F., Zhu, J., He, L., Shao, B., Zheng, N., and Liu, T.-Y.
\newblock Co-evolution transformer for protein contact prediction.
\newblock \emph{Advances in Neural Information Processing Systems},
  34:\penalty0 14252--14263, 2021.

\bibitem[Zhang et~al.(2022)Zhang, Bi, Liang, Cheng, Hong, Deng, Lian, Zhang,
  and Chen]{zhang2022ontoprotein}
Zhang, N., Bi, Z., Liang, X., Cheng, S., Hong, H., Deng, S., Lian, J., Zhang,
  Q., and Chen, H.
\newblock Ontoprotein: Protein pretraining with gene ontology embedding.
\newblock \emph{arXiv preprint arXiv:2201.11147}, 2022.

\end{thebibliography}
